\documentclass[a4paper,12pt]{article}

\usepackage{jheppub} 

\usepackage[T1]{fontenc} 

\usepackage[percent]{overpic}
\usepackage{wrapfig}
\usepackage{bbm}
\usepackage{tabu}
\usepackage{slashed}
\usepackage{diagbox}

\usepackage{graphicx}
\usepackage{tikz}
\usetikzlibrary{arrows,chains,shapes,matrix,positioning,scopes}
\usetikzlibrary{decorations.markings}
\tikzstyle arrowstyle=[scale=1]
\tikzstyle directed=[postaction={decorate,decoration={markings,
    mark=at position .65 with {\arrow[arrowstyle]{stealth}}}}]
\tikzstyle reverse directed=[postaction={decorate,decoration={markings,
    mark=at position .65 with {\arrowreversed[arrowstyle]{stealth};}}}]
\usepackage{import}
\usepackage{accents}
\usepackage{mathrsfs,amsmath,amssymb,slashed}
\usepackage{multirow,multicol}
\usepackage{enumitem}
\usepackage[percent]{overpic}
\usepackage{slashed}

\usepackage{wrapfig}
\usepackage{tabu}
\usepackage{diagbox}
\usepackage{comment}
\usepackage{tikz}
\usepackage{mathrsfs,amsmath,amssymb,amsthm,amsfonts,graphicx,accents,hyperref,color}
\usepackage{leftidx}
\usepackage{import}
\usetikzlibrary{decorations.pathmorphing}
\DeclareFontFamily{OT1}{rsfs}{}

\DeclareFontShape{OT1}{rsfs}{m}{n}{ <-7> rsfs5 <7-10> rsfs7 <10->rsfs10}{} 

\DeclareMathAlphabet{\mycal}{OT1}{rsfs}{m}{n}
\newcommand{\bab}{\boldsymbol{\nabla}}
\newcommand{\tcr}[1]{{\color{red}{#1}}}

\newcommand{\Laa}{\Lambda}

\newcommand{\mone}{{\scriptscriptstyle-1}}
\newcommand{\B}{\mathcal{B}}

\newcommand{\ord}[1]{\mathcal{O}(#1)}
\newcommand{\laa}{\lambda}

\newcommand{\de}{\text{d}}

\newcommand{\A}{\mathcal{A}}
\newcommand{\E}{\mathcal{E}}

\newcommand{\F}{\mathcal{F}}

\newcommand{\Ti}{{\scriptstyle T}}

\newcommand{\nn}{\nonumber}

\newcommand{\be}{\begin{equation}}
\newcommand{\ee}{\end{equation}}

\newcommand\vnote[1]{\textcolor{magenta}{\bf [Vahid:\,#1]}}
\newcommand\enote[1]{\textcolor{red}{\bf [Erfan:\,#1]}}

\newcommand{\df}{\mathbbmss{d}}

\newcommand{\dl}{\mathbbmss{L}}
\title{\centerline{\Large{Source and Response Soft Charges for Maxwell Theory on AdS$_d$}}}
\author{Erfan Esmaeili$^a$,}
\author{Vahid Hosseinzadeh$^a$}
\author{and M. M. Sheikh-Jabbari$^{a,b}$}

\affiliation{\it $^a$ School of Physics, Institute for Research in Fundamental
Sciences (IPM),\\ P.O.Box 19395-5531, Tehran, Iran}

\affiliation{\it $^b$ The Abdus Salam ICTP, Strada Costiera 11, 34151, Trieste, Italy}

\emailAdd{erfanili@ipm.ir}
\emailAdd{v.hosseinzadeh@ipm.ir}
\emailAdd{jabbari@theory.ipm.ac.ir}

\preprint{IPM/P-2019/027}

\abstract{We study asymptotic symmetries and their associated charges for Maxwell theory on anti de Sitter (AdS) background in any dimension. This is obtained by constructing a conserved symplectic structure for the bulk and a theory on the boundary, which we specify. We show that the boundary phase space is described by two scalars and two sets of ``source'' and ``response'' boundary gauge transformations. The bulk dynamics is invariant under these two sets of boundary transformations. We study the (soft) charges associated with these two sets and show that they form an infinite dimensional Heisenberg type algebra. Studying the large AdS radius flat space limit, we show only the source soft charges survive. We also analyze algebra of charges associated with $SO(d-1,2)$ isometries of the background AdS$_d$ space and study how they act on our source and response charges. We briefly discuss implication of our results for the AdS/CFT.}  
\begin{document}
\maketitle
\newpage
\section{Introduction}

{Asymptotic symmetries associated with large gauge transformations have recently  been under intense study. They may be used to label soft modes, which on their own turn, can be used to reformulate soft photon or soft graviton theorems in terms of (spontaneously broken) symmetries, providing a different view on the Ward identities, see e.g. \cite{Strominger:2017zoo, He:2014cra}. The IR dynamics of  gauge theories is relevant to the memory effect \cite{Strominger:2014pwa} and the soft modes may also be relevant to resolving black hole microstate issue \cite{Hawking:2016msc, Afshar:2016uax}.}

{Except for the 3d examples, most of the asymptotic symmetry and soft mode analyses so far has been limited to asymptotic flat spacetimes, see however \cite{Ashtekar:1999jx,Mishra:2017zan, Mishra:2018axf, Compere:2019bua,Anninos:2010zf,Ashtekar:2014zfa,Poole:2018koa,Compere:2008us}. Field theories on the AdS space are of great interest due to the AdS/CFT duality \cite{Aharony:1999ti}. It is hence important to extend the AdS$_3$ analysis to higher dimensions. Dynamics of fields on AdS$_d$, $d>3$ dimensions generically show features not seen at $d=3$ case: Einstein gravity on AdS$_3$ has no propagating degrees of freedom which makes possible some different choices for the boundary conditions \cite{Grumiller:2016pqb}. Maxwell theory on the AdS$_3$ while has propagating photon, allows for electric charges but the corresponding Coulomb  field blows up logarithmically at the AdS boundary.  The question we tackle in this work is studying asymptotic symmetries of Maxwell theory on AdS$_d$, $d\geq 4$.}

{Dynamics of fields on the AdS space is different from the flat space because of its different causal structure and not being globally hyperbolic \cite{Hawking:1973uf}. 
{Nonetheless, one may still formulate ``initial+boundary'' value problem \cite{Ishibashi:2004wx}.} Moreover, as the AdS/CFT has taught us, one may focus more on the  ``radial evolution on AdS'' by viewing the e.o.m of fields on the AdS as evolution equations along the radial direction $\rho$ of the AdS space.\footnote{This AdS radial evolution is  nothing but the holographic renormalization \cite{Skenderis:2002wp}; the Callan-Symanzik and RG flow equations from the dual boundary CFT viewpoint.} Values of the fields in the bulk, with a given consistent initial data set at a constant time slice, are then uniquely specified through their values at the boundary $B$ which in our coordinates is sitting at $\rho\to\infty$. $B$ is a timelike and non-compact surface. Depending on the coordinates adopted on AdS$_d$, $B$ can be different manifolds; e.g. for Poincare patch AdS it is $d-1$ dimensional Minkowski space, for global AdS, it is $R\times S^{d-2}$ and for the coordinates we will be adopting it is a dS$_{d-1}$.}

{In this ``AdS radial evolution'', as the usual practice in the AdS holographic renormalization and to set our initial+boundary value problem, we view $B$  a part of AdS spacetime and impose boundary conditions on the fields at the AdS boundary. For this we need to make sure that the flow of bulk energy momentum tensor of each field vanishes at the causal boundary $B$. Thus, we are  left with the  usual normalizable modes in the bulk \cite{Skenderis:2002wp,Aharony:1999ti}. Specifically, we usually (not necessarily) impose the completely reflective Dirichlet boundary conditions for the fields at the boundary. Noting that massless fields/states can reach the AdS boundary at finite coordinate time, this boundary condition hence prevents energy-momentum leaking out through the boundary $B$.}

{Except for the scalars, masslessness of any other massless field is associated with/protected by some symmetries. For fermions this is the chiral symmetry which is a global symmetry. For $p$-forms $(p\geq 1)$ this is usual local symmetry symmetry generated by a $(p-1)$-form and for gravity (spin-2 field) this is the local Poincare symmetry. The usual metric formulation of gravity makes only the diffeomorphism part of the local Poincare symmetry manifest. In this work we will focus on the $u(1)$ one-form theory, i.e. the Maxwell theory, on the AdS$_d$ and study more closely this gauge symmetry within the ``AdS radial dynamics'' discussed above. To study this radial evolution, we introduce boundary data, which involves a scalar $\Phi$ and a vector A$_\mu$ on the boundary. The value of $A_\mu$ is fixed by the (reflective) boundary conditions on the radiation fields in the bulk.}

{ While the reflective boundary conditions on the AdS boundary are invariant under the bulk and boundary gauge transformations, along with the boundary conditions, as in usual AdS/CFT treatments, one can conveniently fix the  ``radial gauge'', setting the radial component of the gauge field ${\cal A}_\rho$ equal to zero. This gauge fixing leaves us with some boundary gauge transformations. The boundary gauge transformations on AdS space, however, show novel different features not shared by their counterparts on the flat space asymptotic region of flat space (either on spatial infinity $i^0$, or past or future null infinity ${\cal I}^\pm$). }

{Our main result in this work is finding two sets of boundary gauge transformations (BGT), which borrowing the usual AdS/CFT terminology, may be conveniently  called source and response transformations, respectively SBGT and RBGT.  These two sets come about in different ways as we explain below. The SBGT is residual part of the bulk gauge transformations (usually called large gauge transformations) after fixing the radial gauge. As such, the scalar field in the boundary  data $\Phi$  transforms  under the SBGT while the vector A$_\mu$, determined by the boundary value of the radiation fields in the bulk, is invariant under BGT. The RBGT, however, are not directly related to the bulk gauge transformations, at least in the radial gauge, which is the most natural, convenient and usual bulk gauge choice on the AdS. To see them, we note that $A_\mu$ may be naturally decomposed  into a ``transverse'' part $\hat A_\mu$ and an exact part, given by a scalar field $\Psi$. The ambiguity/freedom in this decomposition is what gives rise to the response boundary gauge transformation RBGT. Therefore, $\Psi$ is an SBGT invariant while it transforms under the RBGT.  We study the theory on the boundary governing these two scalars $\Phi, \Psi$, the charges associated with the SBGT and RBGT, and the charge algebra. We find that while the set of SBGT (and RGBT) charges commute among themselves, these two sets do not commute and form an infinite dimensional Heisenberg algebra.} 

The decomposition of the boundary gauge field and introduction of the $\Psi$ field discussed above, naturally leads to a boundary action governing the boundary fields $\Phi, \Psi$. To this action, one may associate an energy momentum tensor which is indeed the boundary part of the charges associated with the AdS isometry group (discussed in section \ref{sec:6}). 

{The rest of this paper is organized as follows. In section \ref{sec:2}, we setup the coordinate system we use for the AdS$_d$ background.  In section \ref{sec:3}, we study equations of motion for the Maxwell's theory on AdS$_d$ and through analyzing some simple solutions we choose our boundary falloff behavior. In section \ref{sec:4}, we make a careful analysis of action principle and symplectic structure of the Maxwell theory on AdS$_d$ and extract a boundary action governing dynamics of boundary fields $\Phi, \Psi$. In section \ref{sec:5}, we present the soft charges associated with the SBGT and RBGT and their algebra. We also introduce and discuss the ``AdS antipodal map'' and its connection to CPT symmetry of our boundary theory. In section \ref{sec:6}, we discuss the AdS$_d$ isometries and the associated conserved charges. We also discuss how the AdS$_d$ isometry charges act on our soft mode phase space. In section \ref{sec:7}, we show how our results on AdS$_d$ and those studied in the literature for Maxwell theory on flat space \cite{Kapec:2014zla, Kapec:2015vwa,  He:2015zea, Campiglia:2017mua, Henneaux:2018gfi, Henneaux:2019yqq, Campoleoni:2019ptc, Esmaeili:2019hom, Giddings:2019ofz} could be related to each other through an AdS (large radius) flat space limit. In particular, we show that only the source boundary gauge transformations and the associated soft charges survive the   limit and the response charges become subdominant and do not appear in the limit. The last section is devoted to discussion and outlook. In three appendices we have gathered some technical details of our analysis.}

{\paragraph{Conventions and Notation.} We will use capital Latin indices  $A,B,C$ for denoting the embedding coordinates of the AdS$_d$ space, $X^A, X^B$ and $A,B=-1,0,\cdots, d-1$. The AdS$_d$ space coordinates will be denoted by $x^a,x^b$, with Latin indices of the beginning of the alphabet, $a,b=0,\cdots, d-1$. The radial coordinate will be denoted by $\rho=x^{d-1}$ and the rest of coordinates, which parametrize the AdS$_d$ boundary, will be denoted by Greek indices, $x^\mu,x^\nu$. We will also need to work with constant time slices. The time coordinate on the AdS and its boundary will be denoted by $\tau=x^0$. The rest of directions on the AdS$_d$ or its boundary will be denoted by $x^i, x^j,\ i,j=1,2,\cdots, d-2$, with Latin indices of the middle of the alphabet. That is, the AdS$_d$ coordinates are $(\rho,\tau, x^i)$ and coordinates at the boundary are $(\tau, x^i)$. }

{We adopt the coordinate system where AdS boundary is at $\rho=\infty$ and denote  the AdS boundary by $B$. The constant time slice at time $\tau$ will be denoted by $\Sigma_\tau$ and its boundary, which is its cross-section with AdS boundary $B$, will be denoted by $\partial\Sigma_\tau$. The latter (for various $\tau$) are hence the usual Cauchy surfaces for the spacetime at the boundary.}

{For the Maxwell gauge field on AdS$_d$ we use ${\cal A}_a$, its radial component will be denoted by ${\cal A}_\rho$ and its projection at the boundary by ${\cal A}_\mu$.}

\section{Anti de Sitter space}\label{sec:2}
The $d$-dimensional anti-de Sitter space is defined as the hyperboloid
\begin{equation}
-X_{\mone}^2+X_a X^a=-\ell^2,\qquad a=0,\cdots,d-1
\end{equation}
embedded in $(d+1)$-dimensional Minkowski space with signature $(--+\cdots+)$. The isometries are generated by
\begin{equation}
iL_{AB}=X_A\partial_B-X_B\partial_A\qquad A,B=-1,0,\cdots,d-1
\end{equation}
which form a $(d-1)$-dimensional conformal algebra $SO(d-1,2)$,
\begin{equation}
-i[L_{AB},L_{CD}]=\eta_{AC}L_{BD}+\eta_{BD}L_{AC}-\eta_{AD}L_{BC}-\eta_{BC}L_{AD}\,.
\end{equation}
A Lorentz subgroup $SO(d-1,1)$  can be identified with generators $L_{ab}$ which preserves $X_a X^a$, so that  the algebra can be cast into the form
\begin{subequations}
\begin{align}
i[L_{\mone a},L_{\mone b}]&=L_{ab},\\
-i[L_{ab},L_{\mone c}]&=\eta_{ac}L_{\mone b}-\eta_{bc} L_{\mone a},\\
-i[L_{ab},L_{ce}]&=\eta_{ac}L_{be}+\eta_{be}L_{ac}-\eta_{ae}L_{bc}-\eta_{bc}L_{ae}.
\end{align}
\end{subequations}
We will call $L_{-1 a}$ the ``AdS-translation'' generators since they reduce to translation vectors near the origin (i.e. in the large $\ell$ flat space limit).  They are
\begin{equation}
iL_{\mone a}=-\sqrt{\ell^2+X_b X^b}\ \partial_a-X_a\partial_{\mone },\qquad X_{\mone }>0
\end{equation}

\subsection{De Sitter slicing of AdS}
We would like to set a coordinate system which makes the Lorentz  $SO(d-1,1)$ symmetry generated by $L_{ab}$ manifest and also allows for taking the flat space limit in a simple way. That is achieved by the parametrization
 \begin{equation}
 X^a X_a\equiv -(X^0)^2+X^iX_i=\rho^2>0,\qquad X_{\mone }^2=\ell^2+\rho^2\,.
 \end{equation}
The AdS$_d$ metric becomes\footnote{
The global AdS coordinates are related to embedding coordinates by
\begin{equation}\label{glo hyper}
    X^0=\sqrt{\ell^2+r^2}\sin t\qquad X^{\mone}=\sqrt{\ell^2+r^2}\cos t\qquad r^2=X^i X_i\nn
\end{equation}
so that the metric is
\begin{equation}
    ds^2=-(\ell^2+r^2)dt^2+\frac{dr^2}{1+r^2/\ell^2}+r^2d\Omega_{d-2}^2
\end{equation}}
\begin{equation}\label{adsmetric}
ds^2=\frac{\ell^2 d\rho^2}{\rho^2+\ell^2}+\rho^2h_{\mu\nu}dx^\mu dx^\nu,\qquad \mu,\nu=0,1,\cdots,d-2\,,
\end{equation}
where $h_{\mu\nu}$ is the $(d-1)$-dimensional Lorentzian metric on unit radius de Sitter space. A convenient choice of coordinate system is \cite{Compere:2017knf}
\begin{equation}\label{dS-metric}
h_{\mu\nu}dx^\mu dx^\nu=\frac{1}{\cos^2\tau}\Big(-d\tau^2+ d\Omega^2_{d-2}\Big),\
\end{equation}
where
\begin{equation}\label{hyp emb}
X^0=\rho\tan\tau,\qquad \sqrt{{X}^i{X}_i}=\frac{\rho}{\cos\tau}\,.
\end{equation}

By the choice of radial coordinate $\rho$, the induced boundary metric will be dS$_{d-1}$, a positively curved, Lorentzian maximally symmetric space. This property facilitates the study of isometries on the AdS$_d$ boundary. Moreover, these coordinates are suitable for taking the flat space $\ell\to\infty$ limit to be discussed in section \ref{sec:7}.

\begin{figure}
    \centering
\begin{tikzpicture}[scale=.6]
\draw[very thick, reverse directed] (-3,5)--(-3,-5);
\draw[very thick,  reverse directed] (3,5)--(3,-5);
\draw[dashed] (-3,3)--(3,-3);
\draw[dashed] (-3,-3)--(3,3);
\draw [purple, thick] (0,3) ellipse (3cm and .3cm);
\draw [purple, thick] (0,-3) ellipse (3cm and .3cm);

\draw [blue, reverse directed] (3,3)..controls (0,0)..(3,-3);
\draw [blue, reverse directed] (3,3)..controls (0.75,0)..(3,-3);
\draw [blue, reverse directed] (3,3)..controls (1.5,0)..(3,-3);
\draw [blue, reverse directed] (3,3)..controls (2,0)..(3,-3);
\draw [blue, reverse directed] (3,3)..controls (2.5,0)..(3,-3);
\draw [blue, reverse directed] (3,3)..controls (2.8,0)..(3,-3);
\draw [blue, reverse directed] (-3,3)..controls (-0,0)..(-3,-3);
\draw [blue, reverse directed] (-3,3)..controls (-0.75,0)..(-3,-3);
\draw [blue, reverse directed] (-3,3)..controls (-1.5,0)..(-3,-3);
\draw [blue, reverse directed] (-3,3)..controls (-2,0)..(-3,-3);
\draw [blue, reverse directed] (-3,3)..controls (-2.5,0)..(-3,-3);
\draw [blue, reverse directed] (-3,3)..controls (-2.8,0)..(-3,-3);
%
%
\node[] at( 3.5,4.5) {$\mathscr{I}$};
\node[] at( -3.5,4.5) {$\mathscr{I}$};
\node[rotate=-90] at( 3.5,0) {$\rho=\infty$};
%
\node[] at( 4.4,3) {$\tau=\pi/2$};
\node[] at( 4.6,-3) {$\tau=-\pi/2$};
\draw [ purple] (0,0)..controls (2,.8)..(3,.8);
\draw [ purple] (0,0)..controls (2,-.8)..(3,-.8);
\draw [ purple] (0,0)..controls (2.1,1.4)..(3,1.4);
\draw [ purple] (0,0)..controls (2.1,-1.4)..(3,-1.4);
\draw [ purple] (0,0)..controls (2.3,1.8)..(3,1.8);
\draw [ purple] (0,0)..controls (2.3,-1.8)..(3,-1.8);
\draw [ purple] (0,0)..controls (2.4,2.1)..(3,2.1);
\draw [ purple] (0,0)..controls (2.4,-2.1)..(3,-2.1);
\draw [ purple] (0,0)..controls (-2.1,1.2)..(-3,1.2);
\draw [ purple] (0,0)..controls (-2.1,-1.2)..(-3,-1.2);
\draw [ purple] (0,0)--(3,0);

\node[purple] at( -3.8,1.2) {{$\Sigma_{\tau}$}};
\node[purple] at( -3.8,-1.2) {${\Sigma_{-\tau}}$};
\end{tikzpicture}
\caption{The patch of AdS$_d$ covered by hyperbolic coordinates. Blue curves are constant $\rho$ hypersurfaces, preserved by $SO(d-1,1)$ Lorentz transformations and the red lines represent codimension one constant time $\tau$ slices, in particular $\Sigma_{\tau},\Sigma_{-\tau}$ are two surfaces at $\tau$ and $-\tau$. Note that in our coordinate system all constant time slices for finite $\tau$ pass through $\rho=0$. The arrows on the boundary and on blue curves show the flow of time $\tau$. In this coordinate system, AdS conformal boundary is a maximally symmetric manifold (dS$_{d-1}$ at $\rho\to\infty$), and  it is conformally invariant under ``AdS-translations''.  In this figure the left and right vertical lines respectively denote the North and South poles of the $S^{d-2}$ at the boundary. The points on the left and right vertical lines together with  $\tau\to -\tau$  are mapped onto each other by the PT transformations, the antipodal map.}
    \label{fig1}
\end{figure}
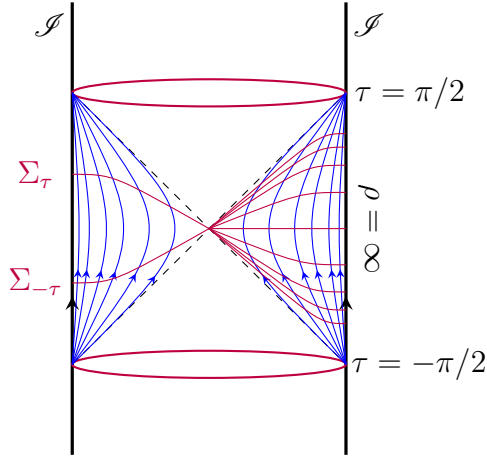

As depicted in Figure \ref{fig1}, there is an AdS PT (parity times time reversal) transformation, $X^A\to -X^A$, which also acts at the AdS boundary as 
$\tau\to -\tau$ and an antipodal map on the $S^{d-2}$. Under this map the radial coordinate $\rho$ does not change. This PT symmetry may be combined with the AdS isometries to form an $O(d-1,2)$ symmetry group.

\subsection{Isometries in the de Sitter slicing}
To the Lorentz generators  $L_{ab}=X_a\partial_b-X_b\partial_a$ one can associate differential 1-forms on the ambient Minkowski space
\begin{equation}
L^{ab}=X^{a}\de X^b-X^b \de X^a\,.
\end{equation}
In the hyperbolic coordinates adapted above they become
\begin{align}
L^{ab}&=(X^{a}\bar{X}^b-X^b \bar{X}^a)\de\rho +\rho^2(\bar{X}^{a}\de\bar{X}^b-\bar{X}^b \de\bar{X}^a)=2\rho^2\bar{X}^{[a} \bar{M}^{b]}_\mu \de x^\mu
\end{align}
where 
\begin{equation}
\bar{X}^a=\frac{X^a}{\rho},\qquad\qquad\bar{M}^{a}_\mu =\frac{\partial \bar{X}^a}{\partial x^\mu}\,.
\end{equation}
Similarly, for the ``AdS-translation'' generators $L_{\mone a}=X_{\mone}\partial_a-X_a\partial_{\mone}$, the corresponding 1-forms are
\begin{align}\label{transl form}
L^{\mone a}&=-X_{\mone} \de X^a+X^a \de X_{\mone}
\end{align}
In hyperbolic coordinates they become
\begin{align}
L^{\mone a}&=(-X_{\mone} \frac{{X}^a}{\rho}+X^a \frac{\rho}{X_{\mone}})\de \rho- X_{\mone}\ \rho \bar{M}^a_\mu \de x^\mu\cr
&=-\frac{\ell^2}{\rho X_{\mone}}X^a\de \rho - \rho X_{\mone}\bar{M}^a_\mu \de x^\mu
\end{align}
The associated vectors using the AdS inverse metric, are
\begin{align}
\text{\textbf{Lorentz:}}\, \xi_L\qquad&\qquad L^{ab}= 2\bar{X}^{[a} \bar{M}^{b]}_\mu h^{\mu\nu}\partial_\nu\label{xi-L} \\
\text{AdS-\textbf{Translation:}}\, \xi_T \qquad&\qquad L^{\mone a}= -X_{\mone}\bar{X}^a\ \partial_\rho - \frac{X_{\mone}}{\rho} \bar{M}^a_\mu h^{\mu\nu}\partial_\nu\label{xi-T}
\end{align}

From \eqref{transl form}, if $\rho\ll\ell$, we have $L^{\mone a}\approx \ell \de X^a$. Hence, $L_{\mone a}$ are translation generators near the origin.
 On the other hand if  $\rho\gg\ell$, i.e. near the boundary of AdS, the AdS-translation generators are
\begin{equation}
   L^{\mone a}\approx  \bar{X}^a\rho\partial_\rho + \bar{M}^a_\mu h^{\mu\nu}\partial_\nu\,.
\end{equation}
Infinitesimal AdS-translations act on $\rho$ as $\rho\to \rho(1+\epsilon \bar{X}^a)$. One can then observe that these
AdS-translations are conformal Killing vectors of the boundary metric in hyperbolic slicing $h_{\mu\nu}$.

The AdS PT transformations, which also act on the dS$_{d-1}$ boundary, may be combined with the Lorentz transformation to form an $O(d-1,1)$ group. As we will discuss in section \ref{sec:5.2}, this PT transformation together with the charge conjugation symmetry gives rise to antipodal map/matching of the boundary theory which acts as a symmetry on our phase space and the corresponding charges.

\section{Physical solutions and boundary conditions}\label{sec:3}

Consider the Maxwell theory on AdS$_d$, described by the action
\begin{equation}\label{Maxwell-action}
S=-\frac14\int d^dx \sqrt{-g} {\cal F}_{ab}{\cal F}^{ab},    
\end{equation}
where ${\cal F}_{ab}=\partial_a {\cal A}_b-\partial_b {\cal A}_a$, $g$ is determinant of AdS$_d$ metric $g_{ab}$ and the indices are raised and lowered by the same metric. To setup the  ``AdS radial evolution'' and describe the ``boundary data'' for the evolution in $\rho$ direction, as discussed in the introduction, we note that equations of motion $\nabla^a {\cal F}_{ab}=0$ involve two classes of equations: those which are first order in $\rho$ and those which are second order. The former may be viewed as constraints among the ``initial'' data set at $\rho=\infty$. These initial data (together with the constraint among them) then ``propagate'' in the radial direction $\rho$ with the latter set of e.o.m. These equations are
\begin{align}
\text{Constraints:}& \qquad D_\mu\F^{\mu\rho}=0\label{heq1}\\
\text{Radial evolution equations:} &\qquad (\rho^2+\ell^2)^{1/2}\rho^{1-d}\partial_\rho\left(\rho^{d-1}(\rho^2+\ell^2)^{-1/2}\F^{\rho \mu}\right)+D_\nu\F^{\nu\mu}=0\label{heq2}
\end{align}
where $D_\mu$ is the covariant derivative with respect to $h_{\mu\nu}$ at constant $\rho$.

\subsection{Charged solutions}\label{Charge Solution Sec}

Any physically relevant boundary condition is expected to allow for moving electric charges. So, we analyze some simple solutions of the field equations  before choosing and setting the boundary conditions.

\paragraph{Boosted electric charge.}
The simplest family of solutions is found by assuming $\A_\mu=\partial_\mu\mathcal{C}$ for which $\F_{\mu\nu}=0$. By closedness of $\F_{ab}$ we have
\begin{equation}
\partial_\mu\F_{\nu\rho}-\partial_\nu\F_{\mu\rho}=0.
\end{equation}
Then, $\F_{\mu\rho}=\partial_\mu\psi$ for a gauge-invariant scalar $\psi$,
\begin{equation}
  \psi=\A_\rho-\partial_\rho \mathcal{C}\,.
\end{equation}
Equation \eqref{heq2} reduces to
\begin{equation}
\frac{\sqrt{\rho^2+\ell^2}}{\rho^{d-1}}\partial_\rho\left(\rho^{d-1}(\rho^2+\ell^2)^{-1/2}\F^{\rho \mu}\right)=0.
\end{equation}
and the solution is (see appendix \ref{Appen:A})
\begin{equation}\label{boosted psi}
\psi=\rho^{3-d}(\ell^2+\rho^2)^{-1/2}\bar{\psi}(x^\nu),\qquad D^\mu D_\mu\bar{\psi}(x^\nu)=0
\end{equation}

This is the exact solution for an electric charge with an arbitrary boost, which crosses the event at the origin of AdS (no dipole moment). To see this, consider a static electric charge at the origin, which in hyperbolic coordinates is given by 
\begin{equation}
 \F_{\tau\rho}=\frac{\rho^{3-d}\cos^{{d-3}}\tau}{\sqrt{\rho^2+\ell^2}}=\partial_\tau\psi(\rho,\tau).  
\end{equation}
$\psi$ is a Lorentz scalar and under boosts,  it acquires  angle-dependence so that other components of $\F_{\mu\rho}$ are also turned on. The phase space of all superpositions of boosted electric charges consists of all functions $\psi(\rho,x^\mu)$ satisfying conditions \eqref{boosted psi}. 
Finally, note that near the origin, $\rho\ll \ell$, the behavior is $\psi\propto \rho^{3-d}$ which matches flat space solutions \cite{Campiglia:2017mua, Esmaeili:2019hom}. Near the boundary, however, the behavior is weaker $\psi\propto \rho^{2-d}$.

\paragraph{Displaced electric charge.}
Let us now relax the condition $\A_\mu=\partial_\mu\mathcal{C}$. Fixing the $\A_\rho=0$ gauge, the constraint equation \eqref{heq1} implies that $\A_\mu$ is a divergenceless de Sitter vector $D^\mu\A_\mu=0$. Spectrum of the Laplace operator on  $\A_\mu$ is given by representation theory of $SO(d-1,1)$ (see appendix \ref{Appen:A} for a derivation using differential equations):
\begin{equation}\label{eigen}
    D^\nu\F^{[\beta]}_{\nu\mu}=-\beta(\beta+d-4)\A^{[\beta]}_\mu\,.
\end{equation}
 Eq.~\eqref{heq2} then becomes an equation for the eigen-vector $\A^{[\beta]}_\mu$:
\begin{equation}\label{eigen eom}
    \frac{\sqrt{\rho^2+\ell^2}}{\ell^2\rho^{d-5}}\partial_\rho\left(\rho^{d-3}(\rho^2+\ell^2)^{1/2}\partial_\rho\A^{[\beta]}_\mu\right)-\beta(\beta+d-4)\A^{[\beta]}_\mu=0.
\end{equation}
 Exact expressions  can be expressed in terms of hypergeometric functions. We are only interested in two asymptotic limits $\rho\ll \ell$ and $\rho\gg\ell$.
\paragraph{Flat region $\rho\ll \ell$.} The two solutions for eigen-vector fields \eqref{eigen} are
\begin{equation}
    \A_\mu^{[\beta]}\propto \rho^\beta,\qquad \qquad \A_\mu^{[\beta]}\propto\rho^{4-d-\beta}\,.
\end{equation}
For integer $\beta$, the solutions with $\rho^{4-d-\beta}$ falloff are of course recognized as the multipole moments of the electromagnetic field \cite{Jackson:1998nia,Esmaeili:2019hom,Herdeiro:2015vaa}. Note that \eqref{eigen} is Lorentz covariant. Therefore, the eigen-vector field $\A_\mu^{[\beta]}$ describes a system of $2^\beta$-poles moving freely in flat space.\footnote{ While non-integer in general, for static  flat space configurations (i.e. electric multipoles discussed in appendix \ref{ APP multi}) $\beta$ becomes a non-negative integer, and by Lorentz covariance of \eqref{eigen}, this remains true for freely moving multipoles. To see this, recall that an electric $2^\beta$-pole configuration in flat space is described by
\begin{equation}
    \A_t=\frac{\mathcal{Y}_{\beta,m_i}}{r^{\beta+d-3}},\nn\end{equation}
where $\mathcal{Y}_{\beta,m_i}$ are harmonics on S$^{d-2}$, \emph{cf.} Appendix \ref{Appen:A} for conventions. In hyperbolic coordinates (which is related to global coordinates by \eqref{glo hyper}), and in radial gauge $\A_\rho=0$ this potential is given by
\begin{equation}
    \A_\tau=\frac{3-\beta-d}{4-\beta-d}\cos^{d+\beta-3}\tau
    \rho^{4-d-\beta}\mathcal{Y}_{\beta,m_i}\qquad\qquad 
    \A_i=-\frac{\tan\tau\cos^{d+\beta-3}\tau}
    {4-\beta-d}\rho^{4-\beta-d}\partial_i\mathcal{Y}_{\beta,m_i}.\nn
\end{equation}
Explicitly, looking for solutions to ${\cal L}_t {\cal A}=0$ in the ${\cal A}_\rho=0$ gauge, we find the above expressions for the fields in which $\beta$ is quantized. Straightforward algebra shows that
\begin{equation}
    D^i\F_{i\tau}=-\beta(\beta+d-4)\A_\tau\,.\nn
\end{equation}
This equality holds also in a boosted frame by Lorentz covariance. Thus, $\A_{\mu}^{[\beta]}$ eigen-vectors with integral values for $\beta$
correspond to electric $2^\beta$-poles. 
}

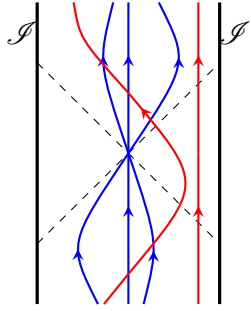
\begin{figure}
    \centering
\begin{tikzpicture}[scale=.4]
\draw[very thick] (-3,5)--(-3,-5);
\draw[very thick] (3,5)--(3,-5);
\draw[dashed] (-3,3)--(3,-3);
\draw[dashed] (-3,-3)--(3,3);

\draw [thick,blue, directed] (0,0)..controls (2,3)..(1,5);
\draw [thick,blue, reverse directed] (0,0)..controls (-2,-3)..(-1,-5);
\draw [thick,blue, directed] (0,0)..controls (-1,3)..(-.5,5);
\draw [thick,blue, reverse directed] (0,0)..controls (1,-3)..(.5,-5);
\draw [directed, blue,thick](0,0)--(0,5);
\draw [directed, blue,thick](0,-5)--(0,0);
\draw [directed, red,thick](2.3,0)--(2.3,5);
\draw [directed, red,thick](2.3,-5)--(2.3,0);
\draw [thick,red, directed] (-.8,-5)--(0,-4)..controls (2.5,-1)..(0,2)..controls (-1.6,3.8)..(-1.8,5);
%
\node[] at( 3.5,4) {$\mathscr{I}$};
\node[] at( -3.5,4) {$\mathscr{I}$};
%
%

\end{tikzpicture}
\caption{Displaced and boosted charges. Blue curves show geodesics that cross the origin at $t=0$. The paths correspond to different Lorentz boosts applied on a static electric charge at the origin. The red curves show two displaced charges, which never cross the origin, or cross it at a later time.}
    \label{fig2}
\end{figure}

\paragraph{Boundary region $\rho\gg \ell$.}
For each eigen-vector field $\A_\mu^{[\beta]}$ one employs a near boundary expansion
\begin{equation}
    \A_\mu^{[\beta]}=\sum\rho^{s}A_\mu^{[\beta](s)}\,.
\end{equation}
Equation \eqref{eigen eom} leads to a recursive relation
\begin{equation}
    s(d+s-3)A^{[\beta](s)}_\mu=\beta(\beta+d-4)A^{[\beta](s+2)}_\mu
\end{equation}
Although the recursive relation depends on the moment $\beta$ (and the exact solution is a hypergeometric function 
depending on $\beta$ and $d$), the leading term is universal in $\beta$
\begin{equation}
    \A^{[\beta]}_\mu\sim\ord{1},\qquad\qquad  \A^{[\beta]}_\mu\sim\ord{\rho^{3-d}}\,\qquad\forall \beta
\end{equation}
In other words, the leading term in the asymptotic series is either $s=0$ or $s=d-3$ for all moments $\beta$. This result reveals the drastic difference between asymptotic behavior of multipole moments in flat and Anti de Sitter spaces. In flat space, higher-poles fall successively faster at far regions. In Anti de Sitter space, however, all multipoles have the same order of magnitude near the boundary. In other words, all the data inside are ``accessible'' to a boundary observer.

\paragraph{Summary of the analysis of charged solutions  at large $\rho$:}
\begin{enumerate}
    \item We will work in radial gauge
   \begin{equation}
        \A_\rho=0\,,\qquad d>3.
    \end{equation}
    This is an unnecessary but convenient choice  which
    makes the formulae simpler and nicer.   
    \item To have multipole configurations, we need to allow
    \begin{equation}
        \A_\mu\sim\ord{\rho^{3-d}}.
    \end{equation}
    \item For $d>3$, the boundary conditions must be relaxed to allow large gauge transformations which have nontrivial charges. The radial gauge leaves boundary gauge transformations with no subleading terms:
\begin{equation}\label{bc 3}
\delta\A_\mu=\partial_\mu\Phi\sim\ord{1},\qquad \partial_\rho\Phi=0.
\end{equation}
\end{enumerate}
As a result, the asymptotic behaviour of the field strength components are,
\begin{equation}\label{F bc}
\F_{\mu\rho}\sim\ord{\rho^{2-d}},\qquad \F_{\mu\nu}\sim\ord{\rho^{3-d}}.
\end{equation}
One may also argue for the above by studying the flux of the energy-momentum tensor, $T^{ab}=\F^{ac}{\F^b}_{c}-\frac14 g^{ab}\F \cdot \F$, 
\begin{align}
    \int_{B}\sqrt{g} T^{\rho a}\xi_a=\int_{B} \sqrt{g}T^{\rho\rho}\xi_\rho+\int_{B} \sqrt{g}T^{\rho \mu}\xi_\mu.
\end{align}
Then from the fact that,
\begin{align}
    T^{\rho\rho}\sim\ord{\rho^{6-2d}},\,\,\,\,\, T^{\rho \mu}\sim\ord{\rho^{3-2d}},
\end{align}
we conclude, 
\begin{align}
    \int_{B} \sqrt{g} T^{\rho a}\xi_a\sim\ord{\rho^{3-d}}.
\end{align}
The lack of flux for isometry charges justifies boundary conditions \eqref{F bc} in $d>3$. The $d=3$ case should be discussed separately.


\subsection{Boundary conditions}\label{sec:3.2}

{First, we note that  different components of the gauge field in the hyperbolic coordinates have the following large $\rho$ expansion,
\begin{equation}
\begin{aligned}
    &\A_{\rho}=\frac{A_\rho}{\rho^{d-2}}+\frac{A^{(3-d)}_\rho}{\rho^{d-3}}+\cdots \\
    &\A_{\mu}=\partial_\mu \Phi+\frac{A_\mu}{\rho^{d-3}}+\frac{A^{(4-d)}_\mu}{\rho^{d-4}}+\cdots
\end{aligned}
\label{Falloffs}
\end{equation}
In a similar manner the gauge parameter $\lambda$ can be expanded asymptotically,
\begin{align}\label{gauge-trans-expansion}
    \lambda=\lambda_S+\frac{\hat\lambda}{\rho^{d-3}}+\cdots\qquad\qquad
\end{align}
As in the usual AdS/CFT \cite{Skenderis:2002wp}, we fix the radial gauge by setting $\A_\rho=0$. This  gauge is accessible by setting  
${A_\rho}=({d-3})\hat\lambda$ and correspondingly for subleading orders, while  the leading gauge parameter $\lambda_S$ is unconstrained. As we see this gauge fixing is not possible in 3d case, as in this case the electric charge yields a logarithmic function of $\rho$; the 3d should be studied separately. Here we focus on $d>3$ case.}

{Let us now focus on the $\A_\mu$ component. The ``constraint equation'' \eqref{heq1} then yields $D^\mu A_\mu=0$ while $\Phi$ remains unconstrained.  Under $\lambda_S$ gauge transformation, $\Phi$ shifts while $A_\mu$ remains intact. The value of $A_\mu$ is fixed by the reflective boundary conditions on the electromagnetic wave in the bulk. Explicitly, the leading term in the bulk electric field at the boundary is proportional to $A_\mu$ and the leading magnetic field is $D_\mu A_\nu-D_\nu A_\mu$.  Nonetheless, one can decompose $A_\mu$ in \eqref{Falloffs} into an exact part and a transverse part $\hat A_\mu$
\begin{equation}\label{A-Psi-hatA}
A_\mu=\ell(\partial_\mu\Psi+\hat A_\mu), \qquad D^\mu\hat A_\mu=0.    
\end{equation}
The factor of $\ell$ is added, for later convenience, to make $\Phi$ and $\Psi$ boundary field to have opposite and equal scaling mass dimension. The $D^\mu A_\mu=0$ relation then yields 
\begin{equation}\label{Psi-eom}
D_\mu D^\mu\Psi=0.
\end{equation}

Being on the dS$_{d-1}$, which allows for harmonic forms (here a 0-form, a scalar), there is a freedom/ambiguity in making the above decomposition. We quantify this ambiguity through $\lambda_R$ ``boundary gauge transformations''. Explicitly, we have two sets of gauge transformations $\lambda=(\lambda_S, \lambda_R)$ such that
\begin{align}\label{Boundary-Data-variation}
    \delta_\lambda \Phi=\lambda_S,\qquad
    \delta_\lambda \Psi=\lambda_{R},\qquad
    \delta_\lambda \hat A_\mu=-\partial_\mu\lambda_R,\qquad D_\mu D^\mu\lambda_R=0,
\end{align}
and hence $\delta_\lambda A_\mu=0$. So, our boundary data are given by $(\Phi, \Psi; \hat A_\mu)$ subject to the above boundary gauge transformations.  At this stage the separation of $A_\mu$ into $\Psi$ and $\hat A_\mu$ parts seems arbitrary. Its virtue will however become clear in the next section when we study the symplectic structure and the boundary charges.}

\section{Action principle and conserved symplectic form}\label{sec:4}

In this section we supplement the Maxwell action \eqref{Maxwell-action} on AdS$_d \ d>3$, by appropriate boundary terms, so that with the falloffs \eqref{Falloffs} we have a well-defined variation principle. Starting from \eqref{Maxwell-action}, we vary the action to get the equations of motion and the symplectic potential,
\begin{align}
    \delta S= \int_{AdS} \sqrt{g} \nabla_a \F^{ab} \delta \A_b+\int_{AdS} \partial_a \theta^a,
\end{align}
with the symplectic potential  $\theta^a=-\sqrt{g}\F^{ab}\delta \A_b$. On the solutions of local equations of motion, $\delta S \approx \int_{B} \theta^\rho$. With the falloffs \eqref{Falloffs} we have,
\begin{align}\label{ActionPrinciple}
    \delta S &\approx -\int_{B} \sqrt{g}\F^{\rho \mu}\delta \A_\mu=(d-3)\int_{B}\sqrt{h}\big(D_\mu\Psi D^\mu\delta \Phi+\hat{A}_\mu  D^\mu \delta\Phi\big)\\
    &= (d-3)\int_{B}\sqrt{h}D^\mu\Big(\Psi D_\mu\delta \Phi+\hat{A}_\mu \delta\Phi\Big)
    -(d-3)\int_{B}\sqrt{h}\Big(\Psi D^\mu D_\mu\delta \Phi+D^\mu \hat{A}_\mu \delta\Phi\Big)\nonumber
\end{align}
where $D_\mu$ is the covariant derivative with respect to $h_{\mu\nu}$ and  $D^\mu=h^{\mu\nu}D_\nu$. Imposing the boundary gauge condition $D^\mu D_\mu\Phi=0$,  the last term vanishes (recall that $\hat A_\mu$ is transverse $D_\mu\hat{A}^\mu=0$ by  definition) and we have 
\begin{equation}\label{kappa-i}
\theta^\rho=\sqrt{h}D^\mu \kappa_\mu,\qquad     \kappa_\mu=(d-3)\Big(\Psi D_\mu\delta \Phi+\hat{A}_\mu \delta\Phi\Big)
\end{equation}
 which guarantees having a well-defined  action principle. 

We conclude that  for a well-posed action principle with the falloff conditions \eqref{Falloffs} we only need to have the following constraints on the boundary data,
\begin{align}
   D^\mu D_\mu\Phi=0,
\end{align}
which in turn constrains $\lambda_S$ to $D^\mu D_\mu\lambda_S=0$. That is, both of our boundary gauge transformations $\lambda_R, \lambda_S$ satisfy  the Laplace equation on the boundary.


\subsection{Conserved symplectic form}
Symplectic density of the theory $\omega^a$ is computed by anti-symmetric variation of the symplectic potential $\theta^a$,  $\omega^a=\delta \theta^a$. The symplectic structure of the theory is obtained by integration of the symplectic density over a Cauchy surface \cite{Hajian:2015eha}. For globally hyperbolic spacetimes, e.g. (asymptotically) flat spaces, we have such Cauchy surfaces. In contrast, the AdS spacetime is not globally hyperbolic and  does not have a foliation with Cauchy slices. Here we choose the constant $\tau$ slices as the foliation slices and define the symplectic structure as, 
\begin{align}\label{Symplectic Structure}
    \Omega^{\text{\tiny bulk}}_\tau
    =\int_{\Sigma_\tau} \sqrt{g}\tau_a \delta\F^{ab}\delta\A_{b}
\end{align}
where $\tau_a=\partial_a \tau$ is the 1-form defining constant $\tau$ slices \cite{Wald:1984rg}. 

The symplectic density satisfies $\partial_a \omega^a=0$ on-shell. By the stokes theorem the symplectic structure \eqref{Symplectic Structure} is then independent of slices if the symplectic flux at the boundary is vanishing. However this is not the case by our boundary conditions, which leads to,
\begin{align}\label{omflux def}
    \Omega^{\text{\tiny bulk}}_{\tau_2}-\Omega^{\text{\tiny bulk}}_{\tau_1}=\int_{B_{12}} \omega^\rho=\int_{B_{12}} \sqrt{h} D_\mu \delta\kappa^\mu+\int_{B_{12}} \omega^{{\text{\tiny flux}}}
\end{align}
where ${B_{12}}$ is a region of the AdS boundary $B$ with two boundary $\partial\Sigma_{\tau_1}$ and $\partial\Sigma_{\tau_2}$, $\kappa^\mu$ is defined in \eqref{kappa-i} and $\omega^{\text{\tiny flux}}=(3-d)\sqrt{h}(\delta\Psi D^\mu D_\mu\delta \Phi+D^\mu \delta\hat{A}_\mu \delta\Phi)$ which vanishes by our boundary gauge condition and the definition of $\hat{A}_\mu$. Since the first term is a total divergence, we can take $\theta_{\text{b'dry}}=-\sqrt{h}\tau_\mu\kappa^\mu$ and construct the conserved symplectic form as,
\begin{align}\label{Symplectic Structure conserved}
    \Omega\equiv\int_{\Sigma_{\tau}}\omega+\oint_{\partial\Sigma_{\tau}}\omega_{\text{b'dry}}=\int_{\Sigma_\tau} \sqrt{g} \tau_a\delta\F^{ab}\delta\A_{b}+(3-d)\oint_{\partial\Sigma_\tau}\sqrt{h}\tau_\mu\Big(\delta\Psi D^\mu\delta \Phi+ \delta\hat{A}^\mu \delta\Phi\Big)
\end{align}
with the conditions $D^2\Phi=0=D^\mu\hat{A}_\mu$. We now have a well-defined ``initial+boundary'' value problem.

Before going further with computation of charges, let us pause and discuss our result \eqref{Symplectic Structure conserved}: While the symplectic form of the original Maxwell theory was not conserved, we made it conserved by adding a boundary piece. This boundary piece was extracted through the decomposition \eqref{Boundary-Data-variation} and in itself may be viewed as symplectic form of a complement gauge theory with dynamical fields $(\Phi,\Psi)$ and the ``external gauge field'' $\hat A_\mu$.
Explicitly, substituting the decomposition $\A_a=\partial_a\Phi+\bar{\A}_a$ into \eqref{Symplectic Structure conserved} and using the equations of motion, we arrive at,
\begin{equation}\label{Omega-total}
    \Omega=\int_{\Sigma_\tau} \sqrt{g} \tau_a\delta\F^{ab}\delta\bar{\A}_{b}+(3-d)\oint_{\partial\Sigma_\tau}\sqrt{h}\ \tau_\mu\Big(\delta\Psi D^\mu\delta \Phi+ \delta \Phi D^\mu\delta\Psi \Big).
\end{equation}
The surface term can be inferred from a boundary action
\begin{equation}\label{bdry-action}
    S_{\text{\tiny b'dry}}=(3-d)\int_{B} \sqrt{h}\ \partial_\mu\Psi\partial^\mu\Phi,\qquad d\neq 3.
\end{equation}
This action is invariant under $SO(d-1,1)$ isometry group of the boundary. The corresponding energy-momentum tensor is given by
\begin{equation}\label{BEM Tensor}
    T^{\text{\tiny b'dry}}_{\mu\nu}=\frac{-2}{\sqrt{h}}\frac{\delta S^{\text{\tiny b'dry}}}{\delta h^{\mu\nu}}=(d-3)\Big[2\partial_{(\mu}\Psi\partial_{\nu)}\Phi-h_{\mu\nu}\partial\Psi\cdot\partial\Phi\Big].
\end{equation}
The role of this tensor will be clarified when we compute the Lorentz charge of the bulk theory.

We note that the boundary action and symplectic form is invariant under the following  scaling symmetry:\footnote{This scaling symmetry appears as a result of the fact that the boundary theory \eqref{bdry-action} is independent of the AdS radius $\ell$. Note that $\Phi, \Psi$ have opposite mass dimension and hence $\Phi\cdot\Psi$ is dimensionless.}
\begin{equation}\label{scaling-transf}
\Phi\to M^{-1} \Phi,\qquad \Psi \to M\Psi.    
\end{equation}
We shall return to the conserved charge of this scaling symmetry in the next section.


\section{Nontrivial (physical) gauge transformations}\label{sec:5}

After addition of appropriate boundary terms which makes the symplectic form conserved, we can find the charges for gauge transformations. It turns out that there are two sets of gauge transformations which have non-vanishing charges, the source and response gauge transformations $(\lambda_S,\lambda_R)$. The algebra is an Abelian one up to a central term between source and response sectors. In this section we also discuss the ``AdS antipodal map'' and the behavior of our charges under the map.

\subsection{Charges and their algebra}
The charge $Q_\lambda$ of a gauge transformation $\lambda$ is the generator of that transformation on the phase space \cite{Lee:1990nz,Brown:1986nw} defined through the symplectic form as,
\begin{equation}
    \delta Q_\lambda=\Omega(\cdot,\delta_\lambda).
\end{equation}
We have two classes of gauge transformations the ``source charge''  denoted by $Q^S_\lambda$ when $\lambda_R=0$ and  the  ``response charge'' denoted by $Q^R_\lambda$ when $\lambda_S=0$. Using \eqref{Boundary-Data-variation} it turns out that,
\begin{align}\label{source-response-charges}
    \begin{split}
        Q^S_\lambda[\Psi]&=\oint_{\partial\Sigma_\tau} \sqrt{h}\tau_\mu\Big(\lambda_S D^\mu\Psi-\Psi D^\mu\lambda_S\Big),\\
        Q^R_\lambda[\Phi]&=\oint_{\partial\Sigma_\tau} \sqrt{h}\tau_\mu\Big(\lambda_R D^\mu\Phi-\Phi D^\mu\lambda_R\Big),
    \end{split}
\end{align}
where in the  above, to simplify the expressions, we have dropped a $d-3$ factor compared to the canonical charge values. $Q^S_\lambda[\Psi],\ Q^R_\lambda[\Phi]$ are both conserved and independent of $\tau$ as a result of our earlier discussions on having a conserved symplectic structure. This latter may also be directly verified given the expressions of the charges above.  To see this, recall that the boundary gauge parameter $\lambda$ and the boundary fields $\Phi, \Psi$ satisfy Laplace equation. This implies that  $dQ^S_\lambda[\Psi]/d\tau,\ dQ^R_\lambda[\Phi]/d\tau$ become exact forms on the S$^{d-2}$ sphere at $\partial\Sigma_\tau$, and hence vanish.

One can then readily compute the algebra of charges:
\begin{align}\label{charges-algebra}
    \begin{split}
        \{Q^S_\lambda, Q^S_\chi\} =0,&\qquad \{Q^R_\lambda, Q^R_\chi\}=0,\\
        \{Q^S_\lambda, Q^R_\chi\}=-\Omega(\delta_{\laa_S},\delta_{\chi_R})&=\oint_{\partial\Sigma_\tau} \sqrt{h}\ \tau_\mu(\lambda D^\mu\chi-\chi D^\mu\lambda).
        \end{split}
\end{align}
As we can see the ``source'' and ``response'' charges do not commute, which justifies the names. 
Using the decomposition and the formulae in the appendix \ref{Appen:A}, we find
\begin{equation}
\{Q^{S, \sigma}_{l,m_i}, Q^{R,\sigma'}_{l',m_i'}\}=2i\ \delta(\sigma\sigma'+1)\ \delta_{l,l'}\delta_{m_i,m_i'},
\end{equation}
where we used an expansion like \eqref{psi-Sphere-decompose} for the gauge parameters and $\sigma,\sigma'=\pm1$ correspond to 
the two solutions \eqref{dS-Laplace-slon}.

The  conserved charge associated with the scaling symmetry \eqref{scaling-transf} is\footnote{Our  definition of $\boldsymbol{\Delta}$ charge differs from its canonical value by a factor of $d-3$, as also our conventions for the charges $Q^R, Q^S$.}
\begin{align}
   \boldsymbol{\Delta}\equiv \oint \sqrt{h}\tau_\mu(\Psi D^\mu \Phi-\Phi D^\mu \Psi),
\end{align} 
with the algebra,
\begin{align}
    \{\boldsymbol{\Delta} ,Q^S_\lambda\}=-  Q^S_\lambda, \qquad
    \{\boldsymbol{\Delta},Q^R_\lambda\}= Q^R_\lambda.
\end{align}

\subsection{AdS antipodal map}\label{sec:5.2} 

\begin{figure}
    \centering
\begin{tikzpicture}[scale=.4]

\draw[very thick] (-3,5)--(-3,-5);
\draw[very thick] (3,5)--(3,-5);
\draw[dashed] (-3,3)--(3,-3);
\draw[dashed] (-3,-3)--(3,3);
\draw [purple, directed](0,3) ellipse (3cm and .3cm);
\draw [purple, reverse directed](0,-3) ellipse (3cm and .3cm);
\node[] at( 3.5,4.5) {$B$};
\node[] at( -3.5,4.5) {$B$};
\node [cyan] at (-3,3) {$\bullet$};
\node [cyan] at (3,-3) {$\bullet$};

%
%
%
%
%
%
%

%

\begin{scope}[shift={(13,-4)}]
\draw[thick] (0,0)--(-4,4)--(0,8)--(4,4)--(0,0);

\node at (0,0) {$\bullet$};
\node at (0,8) {$\bullet$};
\node at (4,4) {$\bullet$};
\node at (-4,4) {$\bullet$};
\draw [purple] (-4,4)..controls (0,0)..(4,4);
\draw [purple] (-4,4)..controls (0,8)..(4,4);
\node at (0,-.6) {$i^-$};
\node at (0,8.6) {$i^+$};
\node at (4.8,4) {$i^0$};
\node at (-4.8,4) {$i^0$};
\node at (2.8,6.4) {$\mathscr{I}^+$};
\node at (2.8,1.6) {$\mathscr{I}^-$};
\node at (-2.8,6.4) {$\mathscr{I}^+$};
\node at (-2.8,1.6) {$\mathscr{I}^-$};

\draw [dashed](4,4) circle (1.2cm);
\draw [dashed] (3.8,5.2)--(7,7);
\draw [dashed] (3.8,2.8)--(7,1);

\draw [purple](11,7) ellipse (3cm and .3cm);
\draw [dashed] (8,7)--(14,1);
\draw [dashed] (14,7)--(8,1);
\draw [purple](11,1) ellipse (3cm and .3cm);
\draw (8,7)..controls (10,4)..(8,1);
\draw (14,7)..controls (12,4)..(14,1);

\node at (11,8) {$\mathscr{I}_+^-$};
\node at (11,0) {$\mathscr{I}_+^-$};
\node [cyan] at (14,1) {$\bullet$};
\node [cyan] at (8,7) {$\bullet$};

\end{scope}


\end{tikzpicture}
\caption{Antipodal matching in  flat and AdS spaces. As depicted in middle and right figures, in Minkowski space the spatial infinity can be represented by a de Sitter space. Antipodal map is a map from past to future boundary of the dS space. The left figure shows AdS space in dS slicing where the blue points are mapped onto each other under antipodal map. The antipodal matching condition is the boundary condition that maps the state/configurations to their CPT conjugate on dS space. }
    \label{Antip}
\end{figure}
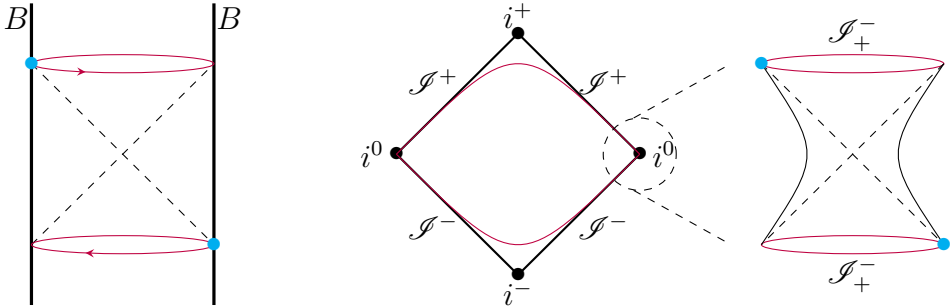

\paragraph{Antipodal map as CPT in flat space.} Local Lorentz invariant quantum field theories in flat space enjoy CPT symmetry. CPT maps \emph{in} states to their CPT conjugate \emph{out} states. On the Penrose diagram of flat space (middle and right diagrams in Figure \ref{Antip}), the \emph{in} and \emph{out} states reside on Cauchy surfaces at $t=\pm T$ for large $T$. All Cauchy surfaces have their boundaries at spatial infinity $i^0$. For physically plausible configurations, the initial massive states are localized at some scale, and the fields revert to vacuum at large enough distances.  Initial massless states are also prepared on a spacelike Cauchy surface, which at large $-T$ tends to $\mathscr{I}^-$.  Thus, \emph{in} states can be prepared on Cauchy surfaces that end on the $t+r\equiv v\to+\infty$ limit of $\mathscr{I}^-$ which is usually denoted by $\mathscr{I}^-_+$. Under PT transformations, $t-r\equiv u\leftrightarrow -v$, so the future of the past null infinity $\mathscr{I}^-_+$ is mapped to the past of the future null infinity $\mathscr{I}^+_-$. The \emph{out} Hilbert space resides on a Cauchy surface at large $T$ which ends at $\mathscr{I}^+_-$.
$\mathscr{I}^+_-$ and $\mathscr{I}^-_+$ are related by PT and fields $\varphi$ are mapped between them by CPT conjugation. 

In the analysis of soft charges, in particular in rederiving Weinberg's soft theorems in the language of soft charges \cite{He:2014cra, He:2015zea, Kapec:2015vwa}, our Hilbert space is the direct product of a transverse (``hard'') and ``soft'' part, ${\cal H}_{tot}={\cal H}_{hard}\otimes {\cal H}_{soft}$. While each of ${\cal H}_{hard}$ and ${\cal H}_{soft}$ parts are invariant under CPT, to recover the soft theorems we need to actually gauge CPT in the ${\cal H}_{soft}$ part. That is, we only keep the CPT even states in the soft Hilbert space. This gauging is called the antipodal matching, for the reason reviewed and described above. 

In our analysis on the AdS (in de Sitter slicing), one may explore similar antipodal map. As depicted in the left Figure in \ref{Antip}, there is a PT symmetry on the AdS space and at its boundary. Moreover, for the fields on AdS or the boundary fields $\Phi, \Psi$ one can consider CPT transformation and hence an antipodal map. The boundary fields and the corresponding gauge transformations are solutions to Laplace equation on de Sitter and hence the two solutions can be written in a basis in which they have definite sign under CPT. (In terms of the $\psi^\pm$ basis introduced in the appendix \ref{Appen:A}, CPT($\psi^\pm$)=$\psi^\mp$; i.e. $\psi^+\pm \psi^-$ are fields with definite sign under CPT.) From the gauge transformation rules \eqref{Boundary-Data-variation}, one can readily see that $\Phi$ (and $\Psi$) field and the corresponding gauge transformation $\lambda_S$ (and $\lambda_R$) should have the same CPT parity. Next, we note  expressions of the charges \eqref{source-response-charges}, which imply that  the charges are non-zero only if the source and response fields ($\Phi, \Psi$) have opposite CPT parity. Therefore, one may introduce an ``AdS-antipodal matching'' in which we keep the even (odd) part of $\Phi$ ($\Psi$) field. However, unlike the flat space case, and as our analysis above shows, this is just a choice and not implied by any physical requirement on AdS.




\section{AdS isometries and their charges}\label{sec:6}
In this section we study the conserved charges associated to the isometry transformations of the AdS background. The boundary conditions, expression of our charges
and the consistency of our phase space relies  on the constraints $D^2\Phi=0=D^\mu\hat{A}_\mu$ which guarantee vanishing of $\omega^{\text{\tiny flux}}$ defined in \eqref{omflux def}. These are all manifestly invariant under Lorentz part of the AdS isometries.The Lorentz charges are hence straightforward to construct. However, the gauge conditions are not covariant under AdS-translations and so the phase space is not consistent with AdS-translations. 
 The resolution to this problem is that behaviour of  the gauge field under AdS-translations must be supplemented by an appropriate (field dependent) gauge transformation, which leaves  the boundary constraints invariant.  

\subsection{\texorpdfstring{$SO(d-1,2)$}{sod} transformations of  boundary data}
For Lorentz transformations labeled by $L_{ab}$ we have (cf. \eqref{xi-L}), 
\begin{align}
    \xi_L=(\xi_L^\rho,\xi_L^\mu)=(0,2\bar{X}^{[a} \bar{M}^{b]}_\nu h^{\mu\nu})
\end{align}
Note that for Lorentz transformations $\xi_L^\rho=0$ and we also have, 
\begin{align}
  D^{}_\mu\xi^L_\nu+D^{}_\nu\xi^L_\mu=0,\qquad D^{}_\nu\xi_L^\nu=0.
\end{align}
The definition $\delta_\xi \A=\mathcal{L}_\xi \A$ implies
\begin{subequations}\label{Loren Lie}
\begin{align}
    &\delta_{\xi_L} \Psi=\xi_L^\nu \partial_\nu \Psi\\
    &\delta_{\xi_L} \Phi=\xi_L^\nu \partial_\nu \Phi\\
    &\delta_{\xi_L} \hat{A}_\mu=\xi_L^\nu D^{}_\nu \hat{A}^{}_\mu+\hat{A}^{}_\nu D^{}_\mu\xi_L^\nu
\end{align}
\end{subequations}
which reflects the fact that they are covariant Lorentz tensors. 

  For AdS-translations labeled by $T^a=L^{\mone a}$ we have (\emph{cf.} \eqref{xi-T}),
\begin{align}
    \xi_T=(\xi_T^\rho,\xi_T^\mu)=(\bar{X}^{\mone}\bar{X}^a\rho,\bar{X}^{\mone} \bar{M}^{a}_\nu h^{\mu\nu}).
\end{align}
AdS-translations satisfy the following relations,
\begin{align}\label{xi-T-iden}
    D^{}_\mu\xi^T_\nu+D^{}_\nu\xi^T_\mu=2\Gamma_{\mu\nu}^\rho \xi^T_\rho=-2\rho\xi_T^\rho h_{\mu\nu},\qquad\qquad
    D_\nu\xi_T^\nu=\frac{1}{\rho}\xi_T^\rho(1-d).
\end{align}
In the $\rho\gg l$ limit only the leading terms are relevant and the expression for the AdS-translation vector simplifies to
\begin{align}
    \xi_T=(\xi_T^\rho,\xi_T^\mu)=(\bar{X}^a\rho,\bar{M}^{a}_\nu h^{\mu\nu}).
\end{align}
It is convenient to define, 
\begin{equation}
    \bar{\xi}^\rho_T=\lim_{\rho\to\infty} \frac{{\xi}^\rho_T}{\rho},\qquad\qquad 
    \bar{\xi}^\nu_T=\lim_{\rho\to\infty}{\xi}^\nu_T.
\end{equation}
A useful implication of Killing equation is the following asymptotic relation
\begin{equation}\label{6.8}
  \bar{\xi}_T^\mu=D^\mu\bar{\xi}^\rho_T\,.
\end{equation}
The boundary fields under AdS-translations transform as\footnote{$\Psi$ appears in the asymptotic expansion of $\A_\mu$ at order $\rho^{3-d}$, \emph{cf.} \eqref{Falloffs} and \eqref{A-Psi-hatA}. Transformation of $\Psi$ is defined such that the spacetime function
$    \Psi^\ast\equiv\frac{\Psi}{\rho^{d-3}}+\cdots$
transforms like a scalar field.
}
\begin{subequations}\label{Translations}
\begin{align}
    \delta_{\xi_T} \Psi&=\bar{\xi}_T^\nu \partial_\nu \Psi+(3-d)\Psi\bar{\xi}_T^\rho,\label{Translations-a}\\
    \delta_{\xi_T} \Phi&=\bar{\xi}_T^\nu \partial_\nu \Phi,\\
    \delta_{\xi_T} \hat{A}_\mu&
    =\bar{\xi}_T^\nu D_\nu \hat{A}_\mu+ \hat{A}_\nu D_\mu\bar{\xi}_T^\nu+(3-d)\bar{\xi}_T^\rho\hat{A}_\mu+(d-3)\Psi\partial_\mu\bar{\xi}_T^\rho.
\end{align}
\end{subequations}


\paragraph{Radial gauge fixing and AdS isometries.} The gauge fixing $\A_\rho=0$ and the expansion 
\begin{align}
    \A_\mu&=\partial_\mu\Phi+\frac{A_\mu}{\rho^{d-3}}+\ord{\rho^{2-d}},\qquad \partial_\rho\Phi=0,\label{rho exp}
\end{align}
are invariant under the Lorentz $SO(d-1,1)$ subgroup of the AdS$_d$ isometries. The AdS-translations, however,  clearly do not  respect these conditions. For example, starting from $\A_\rho=0$,
\begin{equation}
    \mathcal{L}_\xi \A_\rho=\A_\mu\partial_\rho\xi^\mu,
\end{equation}
which is non-vanishing for AdS-translations. In addition, action of AdS-translations on $\Phi$ produces subleading terms which are possibly larger than $\rho^{3-d}$ (actually they are $\ord{\rho^{-2}}$). Both problems are overcome if we supplement AdS-translations by a field-dependent gauge transformation
\begin{equation}
    \A_a\to\A_a+\partial_a\gamma,\qquad \partial_\rho\gamma=-\A_\mu\partial_\rho\xi^\mu.
\end{equation}
This transformation is necessary and sufficient to preserve $\partial_\rho\Phi=0$. As a result, the structure \eqref{rho exp} is respected by all isometry transformations. For AdS-transformations, $\partial_\rho\xi^\mu\sim\ord{\rho^{-3}}$, therefore, $\gamma\sim\ord{\rho^{-2}}$ and hence do not contribute to the  canonical charges we computed in the previous section or to the AdS-isometry charges to be discussed in the next subsection. 

\subsection{Boundary gauge transformations and integrability of AdS isometry charges} 
The above $\gamma$ gauge transformation, however, is not enough, as our other gauge conditions $D^2\Phi=0$ and $D_\nu\hat{A}^\nu=0$ are not covariant under AdS-translations either. While our gauge conditions are Lorentz scalars,
\begin{subequations}
\begin{align}
    &\delta_{\xi_L}(D_\nu D^\nu\Phi)=\mathcal{L}_{\xi_L}(D_\nu D^\nu\Phi)\\
     &\delta_{\xi_L}(D_\nu\hat{A}^\nu)=\mathcal{L}_{\xi_L}(D_\nu\hat{A}^\nu),
\end{align}
\end{subequations}
under AdS-translations they transform as,
\begin{subequations}\label{Translation of gauge}
\begin{align}
    &\delta_{\xi_T}(D_\nu D^\nu\Phi)=D_\nu[\bar{\xi}_T^\nu D_\mu D^\mu \Phi+(d-3)\bar{\xi}_T^\rho D^\nu \Phi]\\
     &\delta_{\xi_T}(D_\nu\hat{A}^\nu)=D_\nu[\bar{\xi}_T^\nu D_\mu\hat{A}^\mu+(d-3)\Psi D^\nu\bar{\xi}_T^\rho].
\end{align}
\end{subequations}

We will introduce other field dependent boundary gauge transformations, which we call $\alpha, \beta$ boundary gauge transformations (BGT),  to preserve AdS-translations too. As we will see, the $\gamma$ transformation above and the $\alpha,\beta$ BGT are of different order in $\rho$ and that the latter contribute to the AdS isometry charges, rendering them integrable.

\subsubsection{\texorpdfstring{$\alpha,\beta$}{alpha-beta} boundary gauge transformations}\label{ALphaBetaSection} 
Equations \eqref{Translation of gauge} show how AdS-translations violate gauge condition on $\Phi$ and are inconsistent with $D_\nu\hat{A}^\nu=0$. This problem can be resolved by supplementing AdS-translations  by an appropriate gauge transformation. The resolution may be provided by {field-dependent} gauge transformations which undo the action of AdS-translations on $\Phi$ and $\Psi$:
\begin{align} \label{alpha beta def}
\delta_\alpha\Phi=-\alpha[\Phi;\xi]\,,\qquad 
    \delta_\beta \Psi=-\beta[\Psi;\xi]\,,\qquad 
    \delta_\beta\hat{A}_\mu=D_\mu\beta[\Psi;\xi], 
\end{align}
where $\alpha$ and $\beta$ specify two classes of boundary gauge transformations  such that
\begin{align}
    (\delta_\alpha+\delta_{\xi_T}) \Phi =0,\qquad 
    (\delta_\beta +\delta_{\xi_T}) \Psi=0.\nn
\end{align}
We denote this phase space action as $\hat\delta_{\xi_T} \equiv \delta_{\alpha\beta}+\delta_{\xi_T}$. 
One can check that the BGCs are invariant under $\hat\delta_{\xi_T}$, \begin{align}
    \hat\delta_{\xi_T} D^\nu D_\nu\Phi=&\delta_{\xi_T} D^\nu D_\nu\Phi-D^2\alpha=0,\nn\\
    \hat\delta_{\xi_T} D^\nu\hat{A}_\nu=&\delta_{{\xi_T}}D^\nu\hat{A}_\nu+D^2 \beta=D_\nu\left[\bar\xi_T^\nu(D^\mu\hat{A}_\mu+D^\mu D_\mu \Psi)\right]\approx 0,\label{divergence variation}
\end{align}
where in the second line we have used the equations of motion.  

The improved AdS-translation of $\hat{A}^\mu$ also would be,
\begin{align}
     \hat\delta_{\xi_T} \hat{A}_\mu &=\delta_{\xi_T} \hat{A}_\mu+D_\mu \beta\\
     &\approx \bar{\xi}_T^\nu D_\nu \hat{A}_\mu+ \hat{A}_\nu
     D_\mu\bar{\xi}_T^\nu+(3-d)\bar{\xi}_T^\rho\hat{A}_\mu+D_\nu\left(\bar{\xi}_T^\nu D_\mu \Psi-\bar{\xi}^T_\mu D^\nu \Psi\right)\\
     &= D_\nu (\bar{\xi}_T^\nu A_\mu-A^\nu \bar{\xi}^T_\mu)
\end{align}
which leads to,
\begin{align}\label{Ahat invariant}
    \hat\delta_{\xi_T}\oint_{\partial\Sigma_\tau} \sqrt{h}\tau_\mu \hat{A}^\mu \approx 0.
\end{align}
This is an interesting conclusion: Consider an electric monopole at the origin. For this configuration $\hat{A}^\mu$ is vanishing and the information of the electric field flux on constant  $\tau$ surfaces (Gauss law) is completely stored in $\Psi$ (see section \ref{Charge Solution Sec}),
\begin{align}
    \oint_{\partial\Sigma_\tau}\sqrt{g}\tau_\mu\F^{\mu\rho}=(d-3)\oint_{\partial\Sigma_\tau}\sqrt{h}\tau_\mu(\hat{A}^\mu+D^\mu \Psi)=(d-3)\oint_{\partial\Sigma_\tau}\sqrt{h}\tau_\mu D^\mu \Psi.
\end{align}
One can translate this monopole from the origin (by the new improved AdS-translations) and construct multipole configurations.\footnote{Note that $\Psi$ and $\hat{A}_\mu$ are covariant under Lorentz transformations, so the argument is unaffected by applying an AdS-Lorentz transformation.} Interestingly, the information of the electric flux is still completely in $\Psi$ although $\hat{A}^\mu$ is not zero and $\hat{A}^\mu$ carries the rest of the information, explicitly,
\begin{align}
    \Psi \longleftrightarrow \text{Monopoles}, \qquad \qquad \hat{A}_\mu \longleftrightarrow \text{Multipoles (excluding the monopole part)}.
\end{align}

\subsubsection{Integrability of isometry charges}  
In this section we show that charges associated with Lorentz transformations $\delta_{\xi_L}$ and  the gauge-supplemented AdS-translation transformations $\hat\delta_{\xi_T}$ are integrable with our boundary conditions. The phase space we are considering consists of configurations of the gauge field $\A_\mu$ respecting our boundary conditions and AdS metric with its all diffeomorphisms. First we observe that if isometry transformations are defined by Lie derivative, none of the isometry charges defined by $\Omega^{\text{\tiny bulk}}$ are integrable, see \eqref{bulk general diff}. However, considering the $\Omega^{\text{\tiny bulk}}+\Omega^{\text{\tiny b'dry}}$ defined in \eqref{Symplectic Structure conserved}, renders the Lorentz charge integrable.  The AdS-translation charges, however, still remain non-integrable. They become integrable once we augment the corresponding diffeos with field-dependent $\alpha,\beta$ gauge transformations. The details of computations is given in the appendix \ref{Append:AdS-charge} and here we mainly present the final results.

\paragraph{Lorentz charges are integrable.}
The action of Lorentz subgroup of isometries defined by Lie derivative enjoys integrable charges. One can see that by computing phase space Lie derivative of the symplectic form under Lorentz transformations, 
\begin{align}
 \hspace*{-3mm}   \dl_{\xi_L}\Omega=\oint \sqrt{h} \tau_\nu \xi_L^\nu \omega^{\text{\tiny flux}},
\end{align}
where $\omega^{\text{\tiny flux}}$ is defined in \eqref{omflux def} and is vanishing on our phase space. Therefore, $\delta_{\xi_L}$ has an integrable charge on the constructed phase space and is computed to be (see \eqref{Lorentz-appendix}), 
\begin{align}\label{Lorentz-charge}
  I_{\xi_L}=\int_{\Sigma_\tau} \sqrt{g} {\tau}_a\xi^L_b T^{ab}+(3-d)\oint_{\partial\Sigma_\tau} \sqrt{h} \tau_\nu \left[\Psi D^\nu(\bar{\xi}_L^\mu D_\mu\Phi)-D^\nu\Psi \bar{\xi}_L^\mu  D_\mu\Phi\right].
\end{align}
The notable point here is the boundary term containing the boundary data $\Psi$ and $\Phi$. 
Using our gauge conditions and Killing equations, one can easily show that the boundary term is nothing but the boundary energy momentum tensor introduced in \eqref{BEM Tensor} contracted by $\xi_L$. So, the Lorentz charges can be recast as,
\begin{align}\label{T-bdry}
 I_{\xi_L}&=\int_{\Sigma_\tau} \sqrt{g} {\tau}_a\xi^L_b T^{ab}+\oint_{\partial\Sigma_\tau} \sqrt{h} \tau_\nu \bar{\xi}^L_\mu T_{\text{\tiny b'dry}}^{\mu\nu}.
\end{align}

\paragraph{AdS-Translation charges are not integrable.} 
Under AdS-translations the symplectic form transforms as (see appendix \ref{Append:AdS-charge} for details), 
\begin{align}\label{STtranslation}
    \dl_{\xi_T}\Omega&=\oint \sqrt{h}\ \tau_\nu \xi_T^\nu \omega^{\text{\tiny flux}}-(d-3)^2\oint \sqrt{h} \ \tau_\nu {\xi}_T^\nu\  \delta\Psi \delta\Phi.
\end{align}
This proves non-integrability of AdS-translations, even if $\omega^{\text{\tiny flux}}=0$.
\paragraph{Improved AdS-translations are integrable.}
In section \ref{ALphaBetaSection} we replaced $\delta_{\xi_T}$ by $\hat\delta_{\xi_T}=\delta_{\xi_T}+\delta_{\alpha\beta}$ which leaves the boundary gauge conditions invariant. 
Detailed computations in appendix \ref{Append:AdS-charge} show that the transformation $\hat\delta_{\xi_T}$ defined above leaves the boundary data $\Psi$ and $\Phi$ invariant and importantly, has integrable well-defined canonical charge. In other words, using the equations of motion we can show that $\hat\delta_{\xi_T}$ leaves our phase space invariant and is also a canonical transformation on it, 
\begin{align}
    \dl_{\hat{\xi}}\Omega=(\dl_{{\xi_T}}+\dl_{{\alpha,\beta}})\Omega \approx 0.
\end{align}
This leads us to,
\begin{align}
    \Omega(\cdot,\hat\delta_{\xi_T})=\delta I_{\hat\xi_T}
\end{align}
where the charge $I_{\hat\xi_T}$  turns out to be (see \eqref{ChargeIVariation}),
\begin{equation}
    I_{\hat\xi_T}=\int_{\Sigma_\tau} \sqrt{g} \tau_a \xi^T_b T^{ab}.
\end{equation}
This is a plausible conclusion since the  boundary data are invariant under the improved AdS-translations.
\subsection{Charge algebra}
One can compute the algebra of canonical charges $I_{\xi_L}$, $I_{\hat\xi_T}$, $Q^S_\lambda$ and $Q^R_\lambda$. First we have,
\begin{align}\label{Lie-T}
 \{I_{\hat\zeta_T},I_{\hat\xi_T}\}&=-\Omega(\hat\delta_{\zeta_T},\hat\delta_{\xi_T})=-\hat\delta_{\zeta_T} I_{\hat\xi_T}\\
    &=-\int_{\Sigma_\tau} \sqrt{g}\tau^b\xi_T^a \delta_{\zeta_T} T_{ab}= -\int_{\Sigma_\tau} \sqrt{g}\tau^b\left([\xi_T,\zeta_T]^a T_{ab}+\mathcal{L}_\zeta( \xi_T^a T_{ab})\right)\\
    &=I_{\widehat{[\xi_T,\zeta_T]}}
\end{align}
where we used the relation
\begin{align}\label{LIE EM TENSOR}
    \mathcal{L}_\zeta( \xi^a T_{ab})=&\zeta^c\nabla_c (\xi^a T_{ab})+ \xi^a T_{ac}\nabla_b \zeta^c
    =\nabla_c(\zeta^c \xi^a T_{ab}-\zeta_b \xi_a T^{ac})
\end{align}
and the fact that the second term in \eqref{Lie-T} is a boundary term and vanishes by the prescribed boundary conditions. 
For the Lorentz case we have the similar situation, 
\begin{align}\label{Lie-L}
 \{I_{\zeta_L},I_{\xi_L}\}&=-\int_{\Sigma_\tau} \sqrt{g}\tau^b\xi_L^a \delta_{\zeta_L} T_{ab}-\oint_{\partial\Sigma_\tau} \sqrt{h} \tau_\nu \bar{\xi}^L_\mu \delta_{\zeta_L} T_{\text{\tiny b'dry}}^{\mu\nu}\\
    &=I_{[\xi_L,\zeta_L]}
\end{align}
where we used the boundary form of relation \eqref{LIE EM TENSOR}. 
The rest Poisson brackets can also easily be computed and the result is,
\begin{subequations}\begin{align}
   \{I_{\hat\xi_T},I_{\hat\zeta_T}\}=I_{\widehat{[\xi_T,\zeta_T]}},\qquad &
\{I_{\zeta_L},I_{\xi_L}\}=I_{[\xi_L,\zeta_L]},\qquad \{I_{\xi_L},I_{\hat\zeta_T}\}=I_{\widehat{[\xi_L,\zeta_T]}}\\
    \{I_{\hat\xi_T},Q^S_\lambda\}&=0, \qquad \qquad
\{I_{\hat\xi_T},Q^R_\lambda\}=0\\ 
    \{I_{\xi_L},Q^S_\lambda\}&=Q^S_{\mathcal{L}_{\xi_L}\lambda}, \qquad 
    \{I_{\xi_L},Q^R_\lambda\}=Q^R_{\mathcal{L}_{\xi_L}\lambda}. 
    \end{align}
\end{subequations}
Among other things, the commutators above show that all our $Q^R, Q^S$ charges commute with the AdS-translation charges, especially with the Hamiltonian and have zero bulk energy. Therefore, they may be called ``soft'' charges.
One may also show that 
\begin{equation}
    \{\boldsymbol{\Delta},I_{\xi_L}\}=0, \qquad
    \{\boldsymbol{\Delta},I_{\hat\xi_T}\}=0. \\ 
\end{equation}

\section{Flat space limit}\label{sec:7}
In this section, by introducing appropriate variables, we show how flat space boundary conditions and charges are recovered at $\ell\to\infty$ limit. First we note that in this limit, the AdS metric \eqref{adsmetric} reduces to
\begin{equation}
ds^2= d\rho^2+\rho^2h_{\mu\nu}dx^\mu dx^\nu,\qquad \mu,\nu=0,1,\cdots,d-2\,,
\end{equation}
which is the Minkowski metric in hyperbolic coordinates \cite{Ashtekar:1991vb}. The $\rho\to\infty$ which is the AdS boundary will also map to spatial infinity of flat space.

In our case of Maxwell theory and working in the radial gauge $\A_\rho=0$, the boundary term of the action is
\begin{equation}
    \delta S\approx \int \rho^{d-1}\sqrt{h}n^\rho g^{\mu\nu}\F_{\rho\mu}\delta\A_\nu= \int \rho^{d-1}\sqrt{h}n^\rho g^{\mu\nu}\partial_\rho\A_{\mu}\delta\A_\nu
\end{equation}
where
\begin{equation}
    n^a=\frac{\sqrt{\rho^2+\ell^2}}{\ell}(1,\mathbf{0}^\mu)
\end{equation}
is the unit normal to the boundary.

Let us define the boundary one-form field $\mathcal{E}$,
\begin{equation}\label{E def}
    \mathcal{E}_\mu\equiv n^\rho \F_{\rho\mu}=
    n^\rho\partial_\rho\A_\mu.
\end{equation}
This would be the ``electric field'' if $\rho$ is considered as the temporal coordinate. Introduction of $\E$ makes the comparison between flat and AdS asymptotics tractable, since the geometric features of AdS space are absorbed in its definition and the boundary term becomes
\begin{equation}\label{bt E}
     \delta S\approx \int \rho^{d-1}\sqrt{h} g^{\mu\nu}\E_\mu\delta\A_\nu\,,
\end{equation}
  and $g^{\mu\nu}$ and $\sqrt{h}$ are the same for flat and AdS metrics. (Non)vanishing of the boundary term depends only on the asymptotic behavior of $\A_\mu$ and $\E_\mu$ and these two are related by definition \eqref{E def}.
 The gauge field $\A_\mu$ can be obtained from $\E_\mu$ up to an integration ``constant'' $\mathscr{A}_\mu\sim\ord{1}$. As a boundary condition we assume that $\mathscr{A}_\mu$ is pure gauge, equal to $\partial_\mu\Phi$ for some function $\Phi$ of boundary coordinates. 
 
 To evaluate \eqref{bt E}, it only remains to determine the behavior of $\E_\mu$. To this end, recall that the equations of motion in the radial evolution \eqref{heq1} and \eqref{heq2} are
 \begin{align}
\text{Constraints:}& \qquad D_\mu\E^{\mu}=0\label{heq-E1}\\
\text{Dynamical equations:} &\qquad \rho^{5-d}n^\rho\partial_\rho\left(\rho^{d-3}\E_{ \nu}\right)+D^\nu\F_{\nu\mu}=0\label{heq-E2}
\end{align}
Taking a further $\rho$-derivative over \eqref{heq-E2} one arrives at a differential equation for $\E_\mu$,
\begin{equation}\label{E eq}
    n^\rho\partial_\rho\left(\rho^{5-d}n^\rho\partial_\rho\left(\rho^{d-3}\E_{ \nu}\right)\right)+D^\nu(\text{d}\E)_{\nu\mu}=0.
\end{equation}
We assume that $\E_\mu$ has a decomposition in terms of eigen-vectors $\E_\mu^{[\kappa]}$
\begin{equation}\label{kappa-sum}
    \E_\mu(\rho,x^\alpha)=\sum_{\kappa=0}f_\kappa(\rho)\ \E_\mu^{[\kappa]}(x^\alpha)
\end{equation}
such that (\emph{cf.} Appendix \ref{Appen:A}),
\begin{equation}\label{E eigen}
    D^\nu(\text{d}\E^{[\kappa]})_{\nu\mu}=-\kappa(\kappa+d-4)\E_\mu^{[\kappa]}\,.
\end{equation}
(Note that here $\kappa$ is not necessarily an integer and is in general a non-negative real number. The sum in \eqref{kappa-sum} may then be viewed as an integral.)

The zeroth component $\E_\mu^{[0]}$ of the electric field satisfies 
 \begin{equation} \label{zero condition}
     D^\nu(\text{d}\E^{[0]})_{\nu\mu}=0\,.
 \end{equation}
Excluding magnetic source configurations of any sort, implies $(\text{d}\E)_{ij}=0$ and   it follows from \eqref{zero condition} that\footnote{
If $(\text{d}\E)_{ij}=0$ then $\E_i=\partial_i Z$. We have
\begin{equation}
    D^i(\text{d}\E^{[0]})_{i\tau}=D^i(\partial_i\E^{[0]}_\tau-\partial_\tau\partial_i Z)=0\quad\Rightarrow\quad \E^{[0]}_\tau-\partial_\tau Z=g^\prime(\tau).
\nn\end{equation}
for some function $g(\tau)$. The claim is proved with $\Psi^\star=Z+g$.}
\begin{equation}
    \E^{[0]}=\partial_\mu\Psi^\star
\end{equation}
All in all, $\E_\nu$ can be decomposed as
\begin{equation}
    \E_\nu=\partial_\nu \Psi^\ast+\E^{[>0]}_\nu\,.
\end{equation}

Equation \eqref{E eq} determines the $\rho$-dependence for each component $\E^{[\kappa]}_\mu$
\begin{equation}
    n^\rho\partial_\rho\left(\rho^{5-d}n^\rho\partial_\rho\left(\rho^{d-3}f_\kappa\right)\right)-\kappa(\kappa+d-4)f_\kappa=0.
\end{equation}

There are two non-trivial solutions for $f_\kappa$ in terms of hypergeometric functions. The one without logarithmic dependence is
\begin{equation}
f_\kappa(\rho)=(\frac{\rho}{\ell})^{3-d-\kappa}\ {}_2F_1[\frac{5-d-\kappa}{2},\frac{-\kappa}{2};\frac{6-d-2\kappa}{2};-\frac{\rho^2}{\ell^2}] 
\end{equation}
We are interested in the two asymptotic  limits:
\begin{align}
\text{AdS\ boundary limit}\qquad   \rho\gg \ell &\qquad\qquad f_\kappa\propto (\frac{\rho}{\ell})^{3-d}\qquad \forall \kappa\geq0\\
\text{Flat space\ limit}\qquad   \rho\ll \ell &\qquad\qquad f_\kappa\propto (\frac{\rho}{\ell})^{3-d-\kappa}
\end{align}
As a result, appropriate  boundary conditions for electromagnetism both in flat and AdS space are suggested as 
\begin{equation}\label{AdSflat bc}
    \E_\nu\sim\ord{\rho^{3-d}}\,.
\end{equation}
In flat space, only the zeroth  component $\E_\mu^{[0]}=\partial_\mu\Psi^\star$ saturates the boundary condition \eqref{AdSflat bc}, while all components in AdS space have the same behavior $\rho^{3-d}$.  Therefore, the electric conserved charges in flat space involve only $\Psi^\star$  \cite{Esmaeili:2019hom}.
 The procedure of making the symplectic form conserved leads to addition of two boundary terms as in \eqref{Symplectic Structure conserved}
\begin{align}\label{Symplectic Structure conserved2}
    \Omega=\int_{\Sigma_\tau} \sqrt{g} \tau_a\delta\F^{ab}\delta\A_{b}+\oint_{\partial\Sigma_\tau}\rho^{d-1}\sqrt{h}\tau^\mu\Big(\delta\Psi^\star D_\mu\delta \Phi+ \delta\E^{[>0]}_\mu \delta\Phi\Big)
\end{align}
This expression is of course equivalent to  \eqref{Symplectic Structure conserved}, if one  reintroduces (recall \eqref{E def})
\begin{equation}
    \Psi^\star=(3-d)\Psi,\qquad \E^{[>0]}_\nu=(3-d)\hat{A}_\nu\,.
\end{equation}
In the flat space $\rho\ll\ell$ limit, however, all $\E_\nu^{[>0]}$ modes become subleading with respect to $\kappa=0$ mode $\Psi^\star$.  As a consequence, the third term in \eqref{Symplectic Structure conserved2} drops out in flat limit and the flat space conserved symplectic form is
\begin{align}\label{Symplectic Structure conserved3}
    \Omega=\int_{\Sigma_\tau} \sqrt{g} \tau_a\delta\F^{ab}\delta\A_{b}+\oint_{\partial\Sigma_\tau}\sqrt{h}\tau^\mu\Big(\delta\Psi^\star D_\mu\delta \Phi\Big)
\end{align}
as proposed in \cite{Campiglia:2017mua, Esmaeili:2019hom}. 

Response transformations \eqref{Boundary-Data-variation}, or equivalently
\begin{equation}
 \delta_\lambda \Psi^\star=\lambda_{R},\qquad
    \delta_\lambda \E^{[>0]}_\mu=-\partial_\mu\lambda_R,\qquad D_\mu D^\mu\lambda_R=0
\end{equation}
exist in anti de Sitter space, since $\Psi^\star\sim \E^{[>0]}_\mu\sim\ord{\rho^{3-d}}$. In flat space, there are no such transformations  due to mismatch of asymptotic behaviours of $\Psi^\star$ and $\E^{[>0]}_\mu$. That is, in flat space we only have source charges and there is no response charge.

\section{Discussion and outlook}\label{sec:8}

We have studied soft charges and large gauge transformations for Maxwell theory on AdS$_d$ and compared the results to the well-known flat space results. For technical reasons and to facilitate taking the flat space limit, we adopted dS slicing of AdS space. We observed that the boundary data may be denoted by two scalar fields $\Phi, \Psi$. This is in contrast to the flat space case where there is {only the gauge transformation associated with $\Phi$ and hence only $Q^S$ soft charges} (see discussions in section \ref{sec:7}). Then, there are two associated boundary gauge transformations $\lambda_S, \lambda_R$ each with their own set of soft charges $Q^S, Q^R$. As another feature of the AdS case unparalleled in the flat case, we noted that all the multipole fields for electrostatic case show the same falloff near the AdS boundary. We also studied the charges associated with the AdS isometries as acting on our boundary phase space labelled by $\Phi,\Psi$. 

Having a well-defined variation principle and a conserved symplectic structure in the bulk  led to a boundary action governing the $\Phi, \Psi$ fields, in which $\dot\Phi$ $(-\dot\Psi)$ is momentum conjugate to $\Psi$ $(\Phi)$. This then led to the algebra of the corresponding soft charges \eqref{charges-algebra}. While we did our analysis in the action (Lagrangian) formulation, it can be instructive to repeat the same analysis in the Hamiltonian formulation, as was done in \cite{Henneaux:2018gfi,Henneaux:2019yqq,Bunster:2018yjr} for Maxwell theory on flat space. It is also desirable to explore the presence of magnetic sources and the role of electromagnetic duality in this context as was done in \cite{Hosseinzadeh:2018dkh,Bunster:2019mup,Andersson:2018juv,Freidel:2018fsk} on flat spacetime.  

As other followup projects, it is desirable to repeat the analysis here for $p$-forms on AdS$_d$. Such analysis in flat space has been carried out in \cite{Afshar:2018apx} (see also \cite{Francia:2018jtb,Campiglia:2018see,Henneaux:2018mgn} for 2-form potentials). It is also interesting to explore  similar problem for gravity theory on AdS$_d,\ d\geq4$ in light of our discussions here, extending the work of Brown-Henneaux \cite{Brown:1986nw} to higher dimensions. There are various arguments and statements in the literature saying that asymptotic symmetry algebra of AdS$_d$ space is just its isometry group $SO(d-1,2)$; see \cite{Safari:2019zmc} for a review and for a proof based on theory of deformations of algebras. 

Finally, one may ask if our analyses and results have any implications for the AdS/CFT. One may note that  the two sides of the gauge/gravity duality enjoy different (local) gauge symmetries. In the gravity side these are diffeomorphisms and on the gauge theory side these are typically some $SU(N)$-type gauge symmetry in the large $N$ limit. The AdS/CFT then relates only the gauge/diffeo invariant (observable) quantities on the two sides. The role of soft charges and their algebra is still an open question in the AdS/CFT setup. Our analysis here suggests that not only a residual part of the bulk gauge symmetry may appear on the boundary ($\lambda_S$), there is a new kind of gauge symmetry also appearing on the boundary $(\lambda_R)$. This latter has no direct counterpart in the bulk, nonetheless, the associated field and (soft) charges $Q^R, Q^S$ are not completely independent and do not commute. Studying further this point can shed a new light on the AdS/CFT and the role of local symmetries in it.

\section*{Acknowledgements}

We would like to thank Gaston Giribet and Kyriakos Papadodimas for discussions and Hamid Afshar, Glenn Barnish, Dario Francia and Ali Seraj for comments on the draft. VH also like to thank the scientific atmosphere of SQS'19 workshop in Yerevan. MMShJ would like to thank the hospitality of ICTP HECAP where this work  finished and acknowledges the support by 
INSF grant No 950124 and Saramadan grant No. ISEF/M/98204.

\appendix

\section{ Spectrum of Laplacian on (A)dS space}\label{Appen:A}
{Our boundary fields and gauge transformations satisfy the Laplace equation on dS$_{d-1}$ space with metric \eqref{dS-metric},
\begin{equation}\label{dS-laplace-massless}
D_\mu D^\mu \psi(x^\nu)=0,
\end{equation}
where $D_\mu$ is the covariant derivative on the de Sitter space. We can decompose $\psi$ in terms of spherical harmonic on S$^{d-2}$, \begin{equation}\label{psi-Sphere-decompose}
\psi=\sum_{l,m_i} \psi_{l,m_i}(\tau) {\cal Y}_{l,m_i},\qquad  \nabla^2_{S^{d-2}} {\cal Y}_{l,m_i}=-l(l+d-3) {\cal Y}_{l,m_i},\quad 
i=1,\cdots,[\frac{d-1}{2}].
\end{equation}
to arrive at
\begin{equation}
 \partial^2_\tau \psi_{l,m_i}+ (d-3)\tan\tau\ \partial_\tau \psi_{l,m_i}+l(l+d-3) \psi_{l,m_i}=0.   
\end{equation}
The solutions are of the form of Gauss' hypergeometric functions $\text{F}=\ _2\!F_1$,
\begin{equation}\label{dS-Laplace-slon}
\hspace*{-3mm}\psi^\pm_{l,m_i}(\tau)={\cal N}_l\ e^{\pm il \tau}\ {\text F}(l,\frac{3-d}{2};l+\frac{d-1}{2} ;-e^{\pm2i\tau}),\quad {\cal N}_l=\frac{2^{\frac{3-d}{2}}}{\Gamma(l+\frac{d-1}{2})}\sqrt{(l-1)!(l+d-3)!}   
\end{equation}
As we can explicitly see $(\psi_{l,m_i}^-)^*=\psi_{l,m_i}^+$ and $\psi_{l,m_i}^+(\tau)=\psi_{l,m_i}^-(-\tau)$. The normalization factor ${\cal N}_l$ has been chosen such that the Wronskian is equal to one,
\begin{equation}\label{dS-Laplace-Wronskian}
\psi^-_{l,m_i}\partial_\tau \psi^+_{l,m_i}-\psi^+_{l,m_i}\partial_\tau \psi^-_{l,m_i} =2i . 
\end{equation}
}
\paragraph{On spectrum of Laplacian on dS$_{d-1}$.} 
For our analysis in section \ref{sec:3} we also need spectrum of Laplacian on dS$_{d-1}$, i.e. 
\begin{equation}\label{dS-laplace-massive}
    D_\mu D^\mu \phi(x^\nu)\equiv D^2 \phi(x^\nu)=\laa \phi(x^\nu),
\end{equation}
which is like the equation of motion of a massive field on dS. Using an expansion like the above in terms of spherical harmonics on S$^{d-2}$ one may reduce \eqref{dS-laplace-massive} to,
$$
 \partial^2_\tau \phi_{l,m_i}+ (d-3)\tan\tau\ \partial_\tau \phi_{l,m_i}+[l(l+d-3)+\frac{\lambda}{\cos^2\tau}] \phi_{l,m_i}=0.   
$$
The solutions of the above are hypergeometric functions whose argument depends on $\lambda$.

Alternatively, and more suitably for our case (recall that we are only interested in large or very small radius expansion of Laplacian on AdS$_d$), we note that dS$_{d-1}$ can be embedded in a $d$-dimensional Minkowski space with coordinates $Y^a$ with metric
$$
ds^2=dr^2+r^2 h_{\mu\nu}dx^\mu dx^\nu,
$$
where $r^2=Y_aY^a$ and $h_{\mu\nu}$ is the metric on dS$_{d-1}$ of unit radius \eqref{dS-metric}. Now, consider the Laplacian on this flat space with dS slicing:
\begin{equation}\label{dS-flat}
    \nabla^2\varphi=\partial^a\partial_a \varphi=r^{1-d}\partial_r\left(r^{d-1}\partial_r \varphi
    \right)
    +\frac{1}{r^2}D^2\varphi=0
\end{equation}
Requiring to be a smooth function at $r=0$ one may expand around $r=0$. Noting that \eqref{dS-flat} is linear in $\varphi$ and is homogeneous in $r$, 
$\varphi=r^\gamma \phi_\gamma(x^\nu)$ where 
\begin{equation}
D^2\phi_\gamma=-\gamma(\gamma+d-2)\phi_\gamma,\qquad \gamma\geq0.
\end{equation}
With solutions,
$$
\phi^\pm_\gamma(\tau)={\cal N}_{l,\gamma}\ \frac{e^{\pm i (l-\gamma)\tau}}{\cos^\gamma\tau}\ {\text F}(l-\gamma,\frac{3-d}{2}-\gamma;l+\frac{d-1}{2} ;-e^{\pm2i\tau}).
$$

For a vector gauge field $A_a$, we consider $\nabla_aF^{ab}=0$ in Minkowski space in the radial gauge $A_r=0$ we then have
\begin{equation}
    r^{1-d}\partial_r\left(r^{d-3}\partial_rA_\mu\right)+\frac{1}{r^4}D^\nu F_{\nu\mu}=0.
\end{equation}
Assuming a regular solution with expansion 
\begin{equation}
    A_\mu=r^\beta \A^{[\beta]}_\mu(x^\nu), \qquad \beta\geq 0
\end{equation}
we have 
\begin{equation}
    D^\nu {\cal F}^{[\beta]}_{\nu\mu}=-\beta(\beta+d-4)\A^{[\beta]}_\mu
\end{equation}


\section{Integrability of AdS charges, computation details}\label{Append:AdS-charge}

In this appendix we  compute  $\Omega(\cdot, \delta_{\xi_L})$ and $\Omega(\cdot, \hat{\delta}_{\xi_T})$ where $\Omega=\Omega^{\text{\tiny bulk}}+\Omega^{\text{\tiny b'dry}}$ and $\hat{\delta}_{\xi_T}=\delta_{\xi_T}+\delta_{\alpha \beta}$. For AdS-translations, there are three  terms to be computed $\Omega(\cdot, \hat{\delta}_{\xi_T})=\Omega^{\text{\tiny bulk}}(\cdot, \delta_{\xi_T})+\Omega^{\text{\tiny b'dry}}(\cdot, \delta_{\xi_T})+\Omega(\cdot,\delta_{\alpha\beta})$. In what follows, we compute each term separately.

\paragraph{Computing $\Omega^{\text{\tiny bulk}}(\cdot,\delta_\xi)$.} Since the bulk symplectic density is a covariant object on this phase space, we have
\begin{equation}\label{bulk general diff}
\Omega^{\text{\tiny bulk}}(\cdot,\delta_\xi)= \delta\int_\Sigma \mathbf{J}_\xi-\oint_{\partial \Sigma}\xi\cdot \boldsymbol{\theta},
\end{equation}
where $\xi$ is a generic diffeomorphism. 
Since $\mathbf{J}_\xi=\boldsymbol{\theta}(\delta_\xi)-\xi \cdot \mathbf{L}$ and $\delta_\xi \A_a=\mathcal{L}_\xi \A_a=\xi^b \F_{ba}+\partial_a(\xi^b \A_b)$, 
\begin{align}
    \int_{\Sigma_\tau} \mathbf{J}_\xi&=\int_{\Sigma_\tau} \sqrt{g} \tau_a (-\xi^c \F^{ab}\F_{cb}-\xi^a \mathcal{L})+\oint_{\Sigma_{\tau \rho}} \sqrt{g} \tau_a\xi^b \A_b \F^{a \rho} \nonumber\\
    &=\int_{\Sigma_\tau} \sqrt{g} \tau_a\xi^c (-\F^{ab}\F_{cb}+\frac14 \delta^a_c \F^{be} \F_{be})+\oint_{\Sigma_{\tau \rho}} \sqrt{g}\tau_a \xi^b \A_b \F^{a \rho} \nonumber\\
    &=\int_{\Sigma_\tau} \sqrt{g} {\tau}_a\xi_b T^{ab}+(d-3)\oint_{\partial\Sigma_\tau} \sqrt{h} \tau_\mu \bar{\xi}^\nu(\hat{A}^\mu D_\nu \Phi+D^\mu\Psi D_\nu \Phi)
\end{align}
where $\Sigma_{\tau\rho}=\partial\Sigma_\tau$ denotes the co-dimension two surface at the constant time slice at large constant $\rho$ and $T_{\mu\nu}$ is the gauge invariant, symmetric energy-momentum tensor of the Maxwell theory. Note that here we have defined,
\begin{equation}
  \boldsymbol{\theta}_{a_1 \cdots a_{d-1}}=\epsilon_{a_1 \cdots a_{d-1}b }\theta^b, \qquad \theta^a=-\sqrt{g}\F^{ab}\delta \A_b, \qquad\qquad \mathbf{L}_{a_1 \cdots a_{d}}=\sqrt{g}L\epsilon_{a_1 \cdots a_{d}},
\end{equation}
where $\epsilon_{a_1 \cdots a_d}$ is the Levi Civita symbol. 

The surface term in \eqref{bulk general diff} is also computed as,
\begin{align}
 \oint_{\partial \Sigma_{\tau}}\xi\cdot \boldsymbol{\theta}=\oint_{\partial \Sigma_{\tau}}\xi^a \theta^b \epsilon_{a b c_1 \cdots c_{d-2} }=\oint_{\partial \Sigma_{\tau}}\tau_\nu(\underbrace{\xi^\nu \theta^\rho}_{\ord{1}}-\underbrace{\xi^\rho\theta^\nu}_{\ord{\rho^{-2}}}).
\end{align}
So, the second term is subleading and dose not contribute, yielding,
\begin{align}
 \oint_{\partial \Sigma_{\tau}}\xi\cdot \boldsymbol{\theta}=(d-3)\oint_{\partial \Sigma_{\tau}}\sqrt{h} \tau_\nu\bar{\xi}^\nu (D_\mu\Psi+\hat{A}_\mu) D^\mu \delta \Phi.
\end{align}
This surface term is not a total variation $\delta Q^{\text{\tiny surface}}$. Therefore, the original bulk symplectic form provides no canonical charge for AdS Killing vector fields.

\paragraph{Computing $\Omega^{\text{\tiny b'dry}}$.}  $\Omega^{\text{\tiny b'dry}}$ is given by the second term in \eqref{Symplectic Structure conserved} and $\Omega^{\text{\tiny b'dry}}(\cdot, \delta_\xi)$ can be easily computed as
\begin{align}\label{total iso var}
   \Omega(\cdot,\delta_{\xi})&=\delta\int_{\Sigma_\tau} \mathbf{J}_{\xi}+(3-d)\oint \sqrt{h} \tau_\nu \left[\delta\Psi D^\nu(\delta_\xi\Phi)-\delta_\xi\Psi D^\nu\delta \Phi\right]\nonumber\\ &+(3-d)\oint \sqrt{h} \tau_\nu ِ\left[(D_\mu\Psi+\hat{A}_\mu)\bar{\xi}^\nu D^\mu \delta \Phi
   +(\delta \hat{A}^\nu\delta_\xi\Phi-\delta_\xi \hat{A}^\nu\delta \Phi)\right].
\end{align}
After a lengthy but straightforward calculation using Killing equations, equations of motion and boundary constraints, we can show that the above formula for $\xi=\xi_L$ leads us to,
\begin{align}\label{Lorentz-appendix}
   \Omega(\cdot,\delta_{\xi_L})=\delta I_{\xi_L}=\delta\int_{\Sigma_\tau} \sqrt{g} {\tau}_a\xi^L_b T^{ab}+(3-d)\delta\oint_{\partial\Sigma_\tau} \sqrt{h} \tau_\nu \left[\Psi D^\nu(\bar{\xi}_L^\mu D_\mu\Phi)-D^\nu\Psi \bar{\xi}_L^\mu  D_\mu\Phi\right].
\end{align}
The above is variation of the Lorentz charge. The first term is variation of the ordinary bulk energy momentum tensor and the second term is a ``boundary energy momentum tensor'' containing our boundary data $\Psi$ and $\Phi$. 
\paragraph{Computing $\Omega(\cdot,\delta_{\alpha\beta})$.} 
For computing the AdS-translations charges we should also compute the charge variation of $\delta_{\alpha\beta}$ boundary gauge transformations. Substituting  $\alpha$ and $\beta$ transformations \eqref{alpha beta def} into the symplectic form we find
\begin{align}\label{Omega-alpha-beta}
 \Omega(\cdot,\delta_{\alpha\beta})=&(3-d)\oint \sqrt{h}\tau_\mu\left(\delta_{\xi_T}\Psi D^\mu\delta\Phi-D^\mu\delta_{\xi_T}\Phi\delta\Psi+\delta_{\xi_T}\Phi D^\mu\delta\Psi-D^\mu\delta_{\xi_T}\Psi\delta\Phi\right).
\end{align}

Adding up \eqref{total iso var} and \eqref{Omega-alpha-beta} we obtain
\begin{align}\label{Omega-total-xi-hat}
        \Omega(\cdot, \hat{\delta}_{\xi_T})&=\delta\int_{\Sigma_\tau} \mathbf{J}_{\xi_T}+(3-d)\oint \sqrt{h} \tau_\nu ِ\left[(D_\mu\Psi+\hat{A}_\mu)\bar\xi_T^\nu D^\mu \delta \Phi   +(\delta \hat{A}^\nu \delta_{\xi_T}\Phi-\delta_{\xi_T}\hat{A}^\nu\delta \Phi)\right]\nn\\
        &+(3-d)\oint \sqrt{h}\tau_\mu \left(\delta_{\xi_T}\Phi D^\mu\delta\Psi-D^\mu\delta_{\xi_T}\Psi\delta\Phi\right)\nn\\
        &=\delta\int_{\Sigma_\tau} \sqrt{g} {\tau}_a\xi^T_b T^{ab}+(3-d)\ {\cal B},
\end{align}
where ${\cal B}$ is the following phase space one-form
\begin{align}
    {\cal B}&=\oint \sqrt{h} \tau_\mu ِ\left[(D_\nu\Psi+\hat{A}_\nu)\bar\xi_T^\mu D^\nu \delta \Phi   +(\delta \hat{A}^\mu\delta_{\xi_T}\Phi-\delta_{\xi_T} \hat{A}^\mu\delta \Phi)\right]\nn\\
        &+\oint \sqrt{h}\tau_\mu \left(\delta_{\xi_T}\Phi D^\mu\delta\Psi-D^\mu\delta_{\xi_T}\Psi\delta\Phi\right)
        -\delta \oint_{\partial\Sigma_\tau} \sqrt{h} \tau_\mu \bar\xi_T^\nu(\hat{A}^\mu D_\nu \Phi+D^\mu\Psi D_\nu \Phi)
\end{align}
If the charge for $\hat{\delta}_{\xi_T}$ is integrable, ${\cal B}$ as a boundary one-form, must be exact on the phase space. One can show that ${\cal B}=0$  by a lengthy but straightforward calculation using Killing equations and equations of motion. We hence conclude that, 
\begin{align}\label{ChargeIVariation}
        \Omega(\cdot, \hat{\delta}_{\xi_T})=\delta I_{\hat\xi_T}=\delta\int_{\Sigma_\tau} \sqrt{g} {\tau}_a\xi^T_b T^{ab}.
\end{align}

\subsection{Integrability of isometries}
Phase space Lie derivative of isometry transformations can be computed straightforwardly, with different results for Lorentz and AdS-translations:
\begin{align}
   \dl_{\xi_L}\Omega&=\delta[\Omega(\delta_{\xi_L},\cdot)]=\oint \sqrt{h} \tau_\nu \bar\xi_L^\nu \omega^{\text{\tiny flux}},\\
    \dl_{\xi_T}\Omega&=\delta[\Omega(\delta_{\xi_T},\cdot)]=\oint \sqrt{h}\ \tau_\nu \bar\xi_T^\nu \omega^{\text{\tiny flux}}-(d-3)^2\oint \sqrt{h} \ \tau_\nu {\bar\xi}_T^\nu\  \delta\Psi \delta\Phi.
\end{align}
where $\omega^{\text{\tiny flux}}$ is defined in \eqref{omflux def}. This proves non-integrability of AdS-translations, even if $\omega^{\text{\tiny flux}}=0$. However, if the isometries are supplemented by field-dependent gauge transformations $\delta_{\alpha,\beta}$, the phase space Lie derivative of the combined trasformations vanishes on-shell
\begin{align}
    \dl_{\hat\xi_T}\Omega=(\dl_{{\xi_T}}+\dl_{{\alpha,\beta}})\Omega\approx 0.
\end{align}

\section{Multipole electric charges in AdS}\label{ APP multi}
As formulated in \cite{Seraj:2016jxi,Compere:2017wrj}, one can associate a conserved Noether charge to each multipole moment of an electrostatic configuration. Restricting ourselves to the static case, we work in the global coordinates $(t,r,x^i)$ where the timelike Killing vector field is manifestly $\partial_t$ and the metric is given as, 
\begin{align}
    ds^2=-(1+\frac{r^2}{\ell^2})dt^2+\frac{1}{(1+\frac{r^2}{\ell^2})}dr^2+ r^2 q_{ij}dx^idx^j.
\end{align}
Working in the static case, the system is completely described by $\A_t$. The relevant equation of motion is,
\begin{align}
    (1+\frac{r^2}{\ell^2})r^{4-d}\partial_r(r^{d-2} \partial_r \A_t) +\Delta\A_t=0
\end{align}
where $\Delta$ is the Laplacian operator on the sphere. Expanding $\A_t$ asymptotically, 
\begin{align}
    \A_t=\sum_n \A_t^{(n)}r^n
\end{align}
and putting into the equation of motion, we find the following recursive relation,
\begin{align}
    -n(d+n-3)\A_t^{(n)}=(n+2)(d+n-1)\A_t^{(n+2)}+\Delta\A_t^{(n+2)}
\end{align}
which terminates only at $n=0$ and $n=3-d$. We take the $n=3-d$ branch of multipole solutions \footnote{The $n=0$ branch of solutions are regular everywhere and are examined in \cite{Herdeiro:2015vaa}.} and put the following boundary conditions, 
\begin{align}
    \A_t=\frac{A_t}{r^{d-3}}+\cdots, \qquad\A_r=0, \qquad \A_i=\partial_i\lambda, \qquad\qquad \lambda(x^m)\sim\ord{1},
\end{align}
where we have used the fact that $\F_{ij}=0=\F_{ir}$ because we do not have magnetic sources and the system is static. 

The symplectic flux is then vanishing by our boundary conditions,
\begin{align}
    \Omega^{{\text{\tiny flux}}}=-\int_B\sqrt{g} \delta \F^{rt}\delta \A_t=0
\end{align}
which is expected from the fact that our system is static and nothing changes from $t$ slice to $t+\delta t$ slice. 
Then the conserved symplectic form is,
\begin{align}
    \Omega=\int_{\Sigma_t} \sqrt{g}\delta \F^{ta}\delta \A_a,
\end{align}
and the conserved charges associated to the large $\ord{1}$ gauge transformation of $\A_i$ is computed as $\Omega(\cdot, \delta_\lambda)=\delta Q_\lambda$ and turns out to be,
\begin{align}
    Q_\lambda=\oint_{\partial\Sigma_t}\sqrt{g}\lambda \F^{tr}=(3-d)\oint_{\partial\Sigma_t}\sqrt{q}\lambda A_t.
\end{align}
We can expand $\lambda$ in terms of the spherical harmonics ${\cal Y}_{l,m}$ on $S^{d-2}$,
\begin{align}
    Q_{lm}=\oint_{\partial\Sigma_t}\sqrt{q}{\cal Y}_{l,m} A_t.
\end{align}
where we ignored the coefficient $(3-d)$. 

As we argued above, multipole configurations in AdS are at the same order compared with monopole.  
This is not the case in flat space case where a $n$-pole has a subleading falloff compared to a $m$-pole for $m<n$. This fact forces us to consider divergent gauge transformations on boundary for being able to see the multipole structure of flat space \cite{Seraj:2016jxi}. In AdS space however, we can see the multipole structure and define the associated conserved multipole charges just by the regular $\ord{1}$ large gauge transformations, as we have done it here. As a final comment, we emphasize the crucial role of staticity in charge conservation in this setup, in contrast to Lorentz invariant treatment presented in this paper.

\bibliographystyle{fullsort.bst}
 
\bibliography{review} 

\end{document}